\documentclass[preprint,12pt]{aastex63}

\newcommand{\msun}{M$_\odot$}

\newcommand{\av}{A_{\rm V}}

\newcommand{\twco}{$^{12}$CO\ }
\newcommand{\twcop}{$^{12}$CO}
\newcommand{\thco}{$^{13}$CO\ }
\newcommand{\thcop}{$^{13}$CO}

\accepted{March 22, 2024}
\shorttitle{GMCs of M31 }
\shortauthors{Lada, Forbrich, Petitpas \& Viaene}
\begin{document}

\title{The Molecular Clouds of M31}

\correspondingauthor{Charles Lada}
\email{clada@cfa.harvard.edu}

\author[0000-0002-0786-7307]{Charles J. Lada}
\affiliation{Center for Astrophysics | Harvard \& Smithsonian, 60 Garden St, MS 72, Cambridge, MA 02138, USA}

\author{Jan Forbrich}
\affiliation{Centre for Astrophysics Research, University of Hertfordshire, College Lane, Hatfield AL10 9AB, UK}
\affiliation{Center for Astrophysics | Harvard \& Smithsonian, 60 Garden St, MS 72, Cambridge, MA 02138, USA}

\author{Glen Petitpas}
\affiliation{Center for Astrophysics | Harvard \& Smithsonian, 60 Garden St, MS 72, Cambridge, MA 02138, USA}
\affiliation{Massachusetts Institute of Technology, Massachusetts Ave, Cambridge, MA 02139, USA}

\author{Sébastien Viaene}
\affiliation{Sterrenkundig Observatorium, Universiteit Gent, Krijgslaan 281, 9000, Gent, Belgium}
\affiliation{Silverfin, Gaston Crommenlaan 12, 9050 Ghent, Belgium}

\begin{abstract}

Deep interferometric observations of CO and dust continuum emission are obtained with the Sub-Millimeter Array (SMA) at 230 GHz to investigate the physical nature of the giant molecular cloud (GMC) population in the Andromeda galaxy (M31).  We use  J = 2-1  $^{12}$CO and $^{13}$CO emission to derive the masses, sizes and velocity dispersions of 162 spatially resolved GMCs. We perform a detailed study of a subset of 117 GMCs that exhibit simple, single component line profile shapes. Examining the Larson scaling relations for these GMCs we find: 1- a highly correlated mass-size relation in both $^{12}$CO and $^{13}$CO emission; 2-  a weakly correlated $^{12}$CO linewidth-size (LWS) relation along with a weaker, almost non-existent, $^{13}$CO LWS relation, suggesting a possible dependence of the LWS relation on spatial scale; and  3-that only 43\% of these GMCs are gravitationally bound. We identify two classes of GMCs based on the strength and extent of  their $^{13}$CO emission.  Examination of  the Larson relations finds that both classes are individually characterized by  strong $^{12}$CO  mass-size relations and much weaker $^{12}$CO and $^{13}$CO LWS relations. The majority (73\%) of strong $^{13}$CO emitting GMCs are found to be gravitationally bound. However, only 25\% of the weak  $^{13}$CO emitting GMCs are bound. The resulting breakdown in the Larson relations in the weak $^{13}$CO emitting population decouples the mass-size and LWS relations demonstrating that independent physical causes are required to understand the origin of each. Finally, in nearly every aspect, the physical properties of the M31 GMCs are found to be very similar to those of local Milky Way clouds.

\end{abstract}

\keywords{Molecular clouds, Andromeda Galaxy, Star formation}

\section{Introduction} 

The recent receiver upgrades of the spectral bandwidth of the Submillimeter Array (SMA) to 32 then 48 GHz enabled the first resolved detections of dust continuum emission from individual Giant Molecular Clouds (GMCs) in an external galaxy (M31) beyond the Magellanic Clouds \citep{2020ApJ...890...42F, v21}. These observations, part of a large multi-year survey of the Andromeda galaxy,  also produced simultaneous and extremely deep measurements of CO and its isotopologues as a consequence of the significant integration times required to obtain the dust continuum detections. The combination of these data provide a level of detailed information regarding the physical natures of molecular clouds in M31 that is only surpassed in studies of GMCs in the Milky Way.  

Our earlier papers dealt with an analysis of the physical cloud properties derived from the dust emission with an emphasis on the measurement of the CO conversion factor or mass-to-light ratio, $\alpha$(CO), for individual GMCs across the disk of M31. 
In this paper we focus on a more detailed analysis of the CO observations obtained in the four seasons of the SMA survey.

At a distance of 781 kpc (McConnachie et al. 2005)  M31, the target of our survey, is the nearest large galaxy to the Sun. 
It is similar in both size and mass to the Milky Way (Kafle et al. 2018; Watkins et al. 2018) and like the Milky Way it is classified as a barred spiral galaxy (Block et al. 2006). However, M31 differs from the Milky Way in a number of ways, some of which could influence the nature of its molecular clouds and the star formation process within them. The molecular mass of M31 is about 1/3 that of the Milky Way (Dame et al. 1993) and the star formation rate is about 1/4 that found in the Milky Way (Tabatabaei \& Berkhuijsen 2010, Elia et al. 2022). Moreover, the structure of the two galaxies differs in ways that might be expected to influence cloud formation and evolution. In particular, M31 has weak spiral structure with most of the molecular gas and star formation contained in a "ring of fire" located between roughly 10 and 14 kpc from the galaxy's center (Koper et al. 1991, Gordon et al. 2006, Fritz et al. 2012). This contrasts with the Milky Way where most of the molecular mass and star formation is found in an inner ring 4 to 7 kpc from the Galactic center where the Milky Way's better defined spiral arms are most tightly wound. The origin of the prominent outer ring structure of M31 has been attributed to a collision with its dwarf companion galaxy, M32,  that happened 210 Myr ago and globally disrupted the structure of its disk (Block et al. 2006; Dierickx et al. 2014).
 Do any of these differences translate into measurable differences with Milky Way clouds on the spatial scales of GMCs? Analysis of our M31 GMC survey was designed, in part, to shed light on this question.

In a seminal paper Larson (1981) identified three relations that are believed to govern the basic properties of Milky Way GMCs. These relations connect the fundamental physical properties of cloud mass, size and velocity dispersion on the spatial scales of the GMCs. As generally formulated Larson's relations consist of two scaling relations, the mass-size and linewidth-size (LWS)  relations and a third relation describing the self-gravitational equilibrium of the clouds. 
The Larson relations have guided studies of Galactic clouds for more than four decades (e.g., Solomon et al 1987, Heyer et al. 2009). However, in external galaxies the relevance of Larson's relations to the nature of the GMC populations within them has not been particularly well established and their very validity even questioned (e.g., Hughes et al. 2013; Colombo et al. 2014). How well do the Larson relations describe the nature of extragalactic cloud populations? Our deep imaging survey of M31 provides us with a unique opportunity  to obtain  measurements of the basic properties (mass, size, velocity dispersion) of individual GMCs in M31 to a degree of precision and robustness exceeded only by studies of clouds in the  Milky Way. Moreover, our observations also provide a unique advantage relative to those of  Milky Way GMCs. From our vantage point  all the clouds in M31 are at the same distance  removing the significant uncertainties in distance that frequently hamper Galactic studies. Because M31 is sufficiently inclined to our line-of-sight we also are not constrained or compromised by having to measure clouds through the mid-plane of a galactic disk as we are in the Milky Way where in most directions inside the solar circle severe overlap of clouds along the line-of-sight degrades our ability to measure accurate cloud sizes and masses.  Moreover, we can probe a much wider area of M31's disk with the same degree of precision, something very difficult to do in the Milky Way given cloud overlaps and distance uncertainties.  In this paper we will investigate the physical nature of the GMC population across M31 and examine in detail the Larson relations for an extragalactic GMC population.
In the first part of the paper we determine the nature of, and the Larson relations for, the general population of GMCs in this galaxy and compare these results to what we know about GMCs in the local Milky Way. In the second part of the paper we identify two physical classes of GMCs that make up the M31 population and compare their physical properties to provide new additional insights into the natures of both the galaxy's GMC population  and the Larson relations.

In section 2 we describe the interferometric observations we will analyze. In section 3.1 we review our data analysis procedures and in section 3.2 discuss some general results. In sections 3.3, 3.4 and 3.5 we investigate the Larson relations between mass and size and line width and size as well as the dynamical state of the clouds, respectively. In section 3.6 we consider whether GMCs are confined by external pressure. In section 3.7 we identify  two classes of GMCs based on the presence of or lack of significant \thco emission and discuss the implications for the Larson relations.  In section 4 we present a summary of our findings and our conclusions.

\section{Observations}

The observations analyzed in this paper were obtained as part of an SMA large program dedicated to a simultaneous survey of 230 GHz continuum and CO line emission from individual molecular clouds across the nearby Andromeda galaxy. A basic description of the receiver and spectrometer configurations and calibration is presented in the two earlier papers in this series (Forbrich et al. 2020, Viaene et al. 2021). The observations reported here consist of 80 individual SMA pointings each acquired with a primary beam (FWHM) of 55 arc sec or $\sim$ 200 pc in the subcompact configuration of the SMA over four consecutive fall seasons (2019-2022). Briefly, this configuration typically uses eight 6 meter antennas and provides a synthesized beam size of $4.5'' \times 3.8''$ at 230 GHz corresponding to a spatial resolution of $\sim 17 \times 14$~pc in M31. The interferometric spatial filtering here provides an optimal resolvable scale of $\lesssim$ 100 pc, making it an excellent match for the  GMCs we are interested in imaging. Occasionally one or two antennas were not in service due to maintenance and/or technical issues in which case the beam size was degraded  to $\sim 8'' \times 5''$ (or  $30 \times 19$~pc). The spectrometer was configured to provide a spectral resolution of 140.0 kHz (i.e., 0.18 km\,s$^{-1}$) per channel. For the first two seasons the instantaneous bandwidth was 32 GHz and for the next two seasons it was upgraded to 48 GHz. These bands were configured to simultaneously cover the frequency range containing the $^{12}$CO, $^{13}$CO and C$^{18}$O $J = 2-1$ emission lines. 
The CO emitting spectral regions were removed from the broad continuum band to enable both line-free measurements of continuum fluxes and separate individual measurements of the three CO isotopologues. In this paper we primarily examine the \twco and \thco measurements. For each isotope a 100~km\,s$^{-1}$ wide band centered in velocity on the peak of the line emission was extracted for the analysis. 

\vfill\eject
\section{Results and Analysis}

\subsection{Sample Definition, and Data Reduction Procedures}

The survey fields examined in this paper consist of a set of SMA pointings toward 80 individual Giant Molecular Associations (GMAs) taken from the catalog of GMAs compiled by Kirk et al. (2015) for this galaxy.  This catalog was constructed from observations of dust emission obtained by the {\sl Herschel} satellite. The sources targeted here are a subset of the Kirk et al. catalog and were selected to cover a range of {\sl Herschel} fluxes with a secondary requirement that the target regions span a large range of galactocentric radii (6-16 kpc) in the disk of M31 so as to ensure, to the extent possible, a set of strong sources that cover a representative range of environmental conditions throughout the disk. Emission from $^{12}$CO and $^{13}$CO was detected in all the surveyed fields.
CO emission from the extracted image cubes was analyzed using custom image and spectral reduction software.  For the CO image analysis in this paper we only examined the line integrated intensity (i.e., moment 0) images. Analysis of the full suite of velocity resolved images contained in the data cubes is deferred to a later paper. However, we did extract spatially integrated CO spectra from the image cubes for each source (cloud) identified in our analysis.  

Initial examination of the moment 0 maps indicated that most fields contained more than one recognizable cloud. To define, identify and extract individual clouds in a \twco or \thco moment 0 image or map we use a watershed segmentation where markers are initially placed at all pixels with values of less than 1 sigma.\footnote{Here sigma is defined as the rms noise  determined from source free regions of each image.} The algorithm then 'floods' the image starting from these markers to determine individual islands defined by the 3 sigma level.  Effectively, we thus isolate individual 3 sigma closed contours. In a final step, we remove pixels that lie outside of the half-power primary beam (r=27.5" or about 100 pc) as well as islands and peaks that are smaller than the synthesized beam size. We define the remaining `closed contour' islands to be clouds only if they are at least partially resolved. We consider a cloud as partially resolved if its area is greater than 1.2 times the area of the synthesized beam. We define the clouds so detected in \twco to be GMCs irrespective of their size or mass.  The closed 3-sigma \twco contour defines the area of the GMC  and we consider any \thco or dust continuum emission within that area to belong to that GMC.  However, since one or more of the extracted \thco clouds can be within the boundaries of a single GMC defined by \twcop, we do not consider the \thco clouds to individually define GMCs and refer to them as clumps. As a result of the segmentation procedure we extracted 162 GMCs in \twco emission and 85 clumps in \thco emission from the 80 observed fields. The 85 \thco clumps were contained in 55 separate GMCs. 

 In a separate step, we then  extract spatially integrated spectra from within the segmented islands.  
  Examples of some of the spectra obtained in this manner can be found in Viaene et al. (2021). For a significant fraction (117/162) of the GMCs the line profiles were found by visual inspection to be well behaved with relatively simple shapes. These profiles were quite effectively fit by single Gaussian functions, although some of these profiles were not strictly pure Gaussian in shape.
  From these fits we obtained the velocity dispersions and integrated CO intensities for the 117 GMCs  along with the corresponding formal uncertainties in those quantities. Those uncertainties were found to be only 3\% and 8\% for the \twco dispersions and integrated intensities, respectively, indicating both the high signal-to-noise and high precision of those spectral observations. The formal uncertainties for \thco spectra were only somewhat higher being 4\% and 15\% for the dispersions and integrated intensities, respectively.   The remaining GMCs were found to have very complex or multiple component line profiles, not at all  well matched by single Gaussian shapes, indicating the possibility of multiple clouds along the line-of-sight to these GMCs. To ensure that our sample consists of the highest possible quality, uncontaminated measurements of GMCs in M31 we perform the bulk of our analysis on the 117 GMCs with relatively clean, single component line profiles. This minimizes introducing significant uncertainties in the derived GMC sizes and masses that would be  invariably associated with ambiguous velocity decomposition and cloud extraction algorithms for any velocity blended clouds.  The relatively large fraction of clouds in M31 that are characterized by simple line profiles compared to that found in Milky Way surveys is likely the result of both our external vantage point and the fact that M31 is sufficiently inclined to our line-of-sight. We refer to the sample containing these 117 GMCs as either the restricted, clean or 1G GMC sample. Forty-four of these GMCs contained 53 of the 85 \thco clumps belonging to the full sample. 

\begin{figure}[t!]
\centering
\includegraphics[width=1.0\hsize]{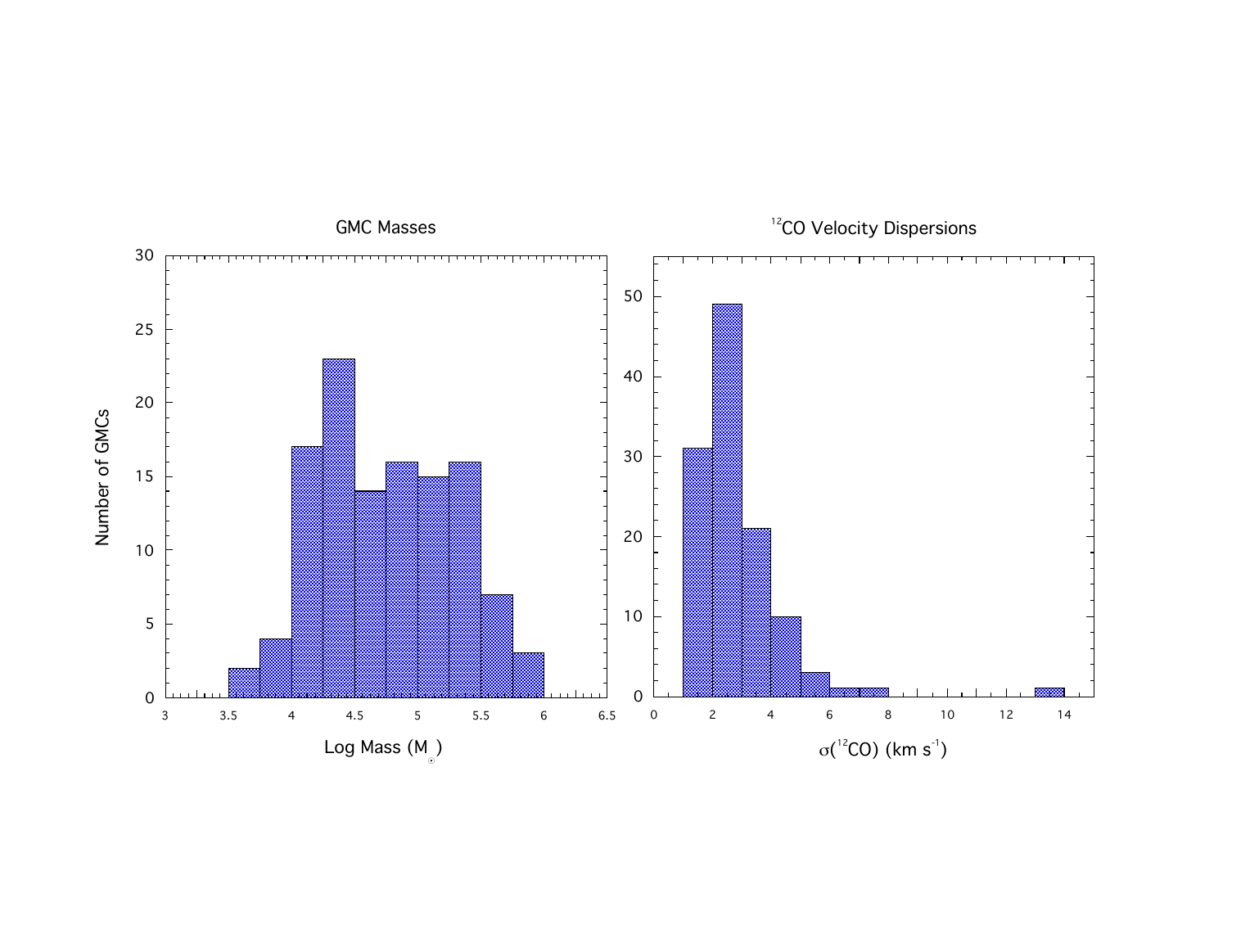}
\vskip -0.8 in
\caption{Basic \twco(2-1) measurements of masses and velocity dispersions in the form of frequency distributions for the clean sample of 117 GMCs in M31 (see text). On the left is the frequency distribution of measured GMC masses and on the right the frequency distribution of velocity dispersions.   \label{MassFn}  }
\end{figure}

\subsection{GMC Masses, Sizes and Velocity Dispersions}

From the segmentation analysis we derive a few basic physical properties for each extracted GMC including the CO integrated intensity, I(CO), and luminosity, L(CO), and the area or effective radius, $R$,  of the cloud. We can then obtain the mass of the GMC from the observed CO luminosity, $L(CO)$, via:
$$M_{GMC} = \alpha_{CO} L(CO)$$
\noindent
where $\alpha_{CO}$ is the dust calibrated CO conversion factor for the total gaseous mass of the cloud, including heavy elements, that is,
$$\alpha_{CO} = DGR \times <\alpha\prime_{CO}\ >$$
\noindent
where $DGR$ is the dust-to-gas ratio, here assumed to be equal to 136 and  $<\alpha\prime_{CO}>$ is the average dust mass-to-light ratio for CO in M31, i.e., $M_{dust}/L(CO)$ where:  
 $$M_{dust} = S_\nu D^2/(\kappa_\nu B_\nu(T))$$
\noindent
where $S_\nu$ is the 230 GHz flux of the dust continuum emission, $D$ the distance to M31, $\kappa_\nu$ the corresponding dust opacity, and $T$ the dust temperature, assumed to be 18~K (Kirk et al. 2015).
Following Viaene et al. (2022) we adopt the dust opacity of $\kappa_\nu =$ 0.0425 from the THEMIS dust model of Jones et al. (2017).  This value is also the same as that empirically derived by Lewis et al. (2022) for local MW clouds at 353 GHz, once adjusted to 230 GHz assuming a power law emissivity index of $\beta =$-2.
The value of  $<\alpha\prime_{CO}>$ in M31 was determined from our earlier observations to be equal to  0.073 ($\pm 0.033$)  for \twco emission and 0.41 ($\pm$ 0.17) for \thco  corresponding to $\alpha_{CO} = 10  \pm 4.5$ and 56 $\pm$ 23 \msun \ (K km s$^{-1}$)$^{-1}$ for \twco and \thco, respectively (Viaene et al. 2022)\footnote{Note that these values adjust those originally listed by Viaene et al. (2022) to account for the slightly lower dust temperature adopted here (i.e, 18 vs. 20 K).}. As mentioned earlier, velocity dispersions for the clouds were derived from Gaussian fits to the spatially integrated CO spectra of each cloud. 
 The formal uncertainties in the masses are relatively small (8\% and 15\% for \twco and \thcop, respectively).  However, these internal uncertainties only partially reflect the total error budget. As will be demonstrated later, that budget is dominated by systematic uncertainties which can be difficult to quantify. 

 We did not attempt to correct the masses for any possible radial metallicity gradient in M31. While studies of M 31 H II regions show a radial metallicity gradient that is relatively flat, about 5-6 times less steep than that measured for the Milky Way, observations of planetary nebulae show no clear evidence for the presence of a radial metallicity gradient in M31  (Bhattacharya et al. 2022, Sanders et al. 2012, Wenger et al. 2019).  Moreover, the scatter in the measured metallicities in these studies is large and, if real, would suggest that spatially random, non-radial metallicity variations may exceed those produced by the relatively flat radial gradient.  So the extent to which one can and should correct GMC masses for possible radial metallicity variations is unclear at the present time.

The basic physical properties of the spatially resolved GMCs and \thco clumps, respectively, are listed in Tables \ref{12codata} and \ref{13codata}.
Figure \ref{MassFn} displays the frequency distributions of masses and velocity dispersions that we measured for the 117 GMCs with single Gaussian component line profiles that constitute our clean sample.
The masses of these clouds range from $3.3 \times  10^3$ to $5.9 \times 10^5$~\msun. The total mass of GMCs in this sample is $1.3 \times 10^7$~\msun. The figure shows that our measurements are uniformly sampled in mass over most of this range. Although the numbers of GMCs are roughly similar in the range between 10$^4$ and $3 \times 10^5$~\msun, the total molecular gas mass contained in the bins is highly skewed toward the high mass bins with slightly more than 80\% of the total mass of this GMC sample found in clouds with M$_{GMC}$ $>$ 10$^5$~\msun. We note here that, although possibly representative,  our survey of M31 GMCs is likely incomplete and consequently the observed mass distribution in the figure very likely does not represent the actual mass function of GMCs in the galaxy. We find the velocity dispersions in this sample to range between 1.3 to 13 km\,s$^{-1}$. However the overall distribution is relatively narrow with 73\% of the values lying between 1.5-3.5 km\,s$^{-1}$ with a dispersion of only 1.5 km\,s$^{-1}$ about the mean of 2.8  km s$^{-1}$. The median value is 2.5 km\, s$^{-1}$ for the sample. Finally we measured the deconvolved radii of the 117 GMCs  to range between 8.5 and 60~pc with an average for the sample of 21 $\pm$ 10 pc.   
 \begin{deluxetable*}{cccccc}
\tablenum{1}
\tablecaption{M31 GMCs : Basic Physical Properties \label{12codata}}
\tablewidth{0pt}
\tablehead{
\colhead{GMC ID}&\colhead{Mass}&\colhead{Radius }&\colhead{Dispersion} &\colhead{Rgal}&\colhead{SGP}\\
\colhead{Kirk\#}&\colhead{M$_\odot$(10$^x$)}&\colhead{pc}&\colhead{km/s}&\colhead{kpc}}
\startdata
K001A&5.93(5)&37.8&4.04&15.2&1G\\
K008B&1.97(5)&17.8&2.30&3.6&1G\\
K008A&4.47(5)&31.9&13.18&3.6&1G\\
K010A&2.67(5)&34.1&3.71&4.8& \\
K022A&5.76(4)&29.6&4.63&3.5&1G\\
K025A&3.33(4)&17.6&2.99&1.4&1G\\
K025B&1.99(4)&14.6&4.80&1.4& \\
K026A&3.92(5)&37.1&2.94&5.8&1G\\
K029A&3.08(5)&31.1&8.76&3.5& \\
K029B&4.88(4)&13.8&7.68&3.5&1G\\
K048A&3.88(5)&34.7&4.04&5.7&1G\\
K056B&8.79(4)&23.0&3.36&9.9& 1G\\
K056A&1.39(5)&26.2&2.89&9.9& \\
K059A&1.04(5)&21.2&2.77&12.5&1G\\
K059B&5.41(4)&17.2&1.91&12.5& \\
K059C&3.07(4)&12.6&2.06&12.5& 							
\enddata
\tablecomments{The source IDs follow the convention that for multiple GMCs within the same SMA field, sources are ordered by mass with "A" referring to the highest mass GMC in the field. SGP refers to sources with single gaussian (1G) line profiles. Only a portion of Table 1 is shown here. The entire table is published in machine readable form in the online version of the paper. }
\end{deluxetable*}

\begin{deluxetable*}{ccccc}
\tablenum{2}
\tablecaption{$^{13}$CO Clumps: Basic Physical Properties\label{13codata}}
\tablewidth{0pt}
\tablehead{
\colhead{Clump ID}&\colhead{Mass}&\colhead{Radius }&\colhead{Dispersion} &\colhead{SGP (\twco)}\\
\colhead{Kirk\#}&\colhead{M$_\odot$(10$^x$)}&\colhead{parsecs}&\colhead{km/s}& }
\startdata
K001Ab &4.05(4)&10.4&2.14& 1G\\
K001Aa&2.79(5)&23.1&3.39&1G\\
K026Aa&3.30(5)&28.5&2.28&1G\\	
K048Aa&2.78(5)&25.6&2.60&1G\\
K056Bb&3.53(4)&14.2&1.46&1G\\
K056Aa&8.87(4)&18.6&2.20&\\
K059Aa&4.24(4)&10.5&2.20& \\		
K063Aa&1.01(5)&20.7&7.55& \\		
K063Bb&1.65(4)&9.0&2.44&1G\\
K067Aa&1.18(5)&20.2&2.98&1G\\	
K071Aa&9.16(4)&15.1&2.26& \\
K071Bb&8.08(4)&13.7&3.11& \\										
\enddata
\tablecomments{The source IDs follow the convention for the $^{12}$CO identified GMCs in Table 1 with the addition of a small letter, a, b, c... to designate the individual $^{13}$CO clumps within the GMC. As with the $^{12}$CO, here the $^{13}$CO sources are ordered by mass with "a" referring to the highest mass clump. The last column lists GMCs with  \twco Single Gaussian Profiles. Only a portion of Table 2 is shown here. The entire table is published in machine readable form in the online version of the paper. }
\end{deluxetable*}

\subsection{The Larson Relations for the Aggregate GMC Population}

\subsubsection{The Mass-Size Scaling Relation} 

The mass-size scaling relation relates the mass of a cloud to its size, usually expressed as a radius, $R= \sqrt{Area/\pi}$. Although, with $R$ derived in this way, the GMC mass-size relation is fundamentally a relation between the mass and area of a population of clouds (e.g., Beaumont et al. 2012), in this paper we will follow the convention of expressing the cloud size in terms of its radius as derived above. Here the cloud masses are the total cloud masses calculated above an outer column density contour set by the 3$\sigma$ noise level in the individual narrow band CO images. Similarly the cloud radii correspond to the total areas encompassing the material above the 3$\sigma$ noise level in an image containing the cloud. 

In his seminal work Larson (1981) found for local clouds a scaling relation between average cloud volume density, i.e., $\rho$ $\equiv$ $3M\over{4\pi R^3}$, and cloud size, $R$, such that $\rho\sim R^{-1}$ or equivalently, $M \sim R^2$, indicating that local Milky Way clouds had nearly constant average surface densities ($\Sigma_{GMC} = M_{GMC}/\pi R_{GMC}^2$). More recent work using dust extinction measurements to trace cloud structure rather than CO found extremely tight correlations between cloud masses and sizes with slopes closely clustered around 2.0 reinforcing the notion of a constant average surface density for local GMCs (Lombardi et al. 2010; Chen et al. 2020; Lewis et al. 2022). The most recent studies of CO in Galactic GMCs have reported mass-size relations with similar but typically somewhat higher slopes than derived from extinction measurements, ranging from 1.9 - 2.4 (Roman-Duval et al. 2010; Miville-Deschenes et al. 2017; Lada \& Dame 2020; Lewis et al. 2022).

\begin{figure}[t!]
\centering
\includegraphics[width=0.6\hsize]{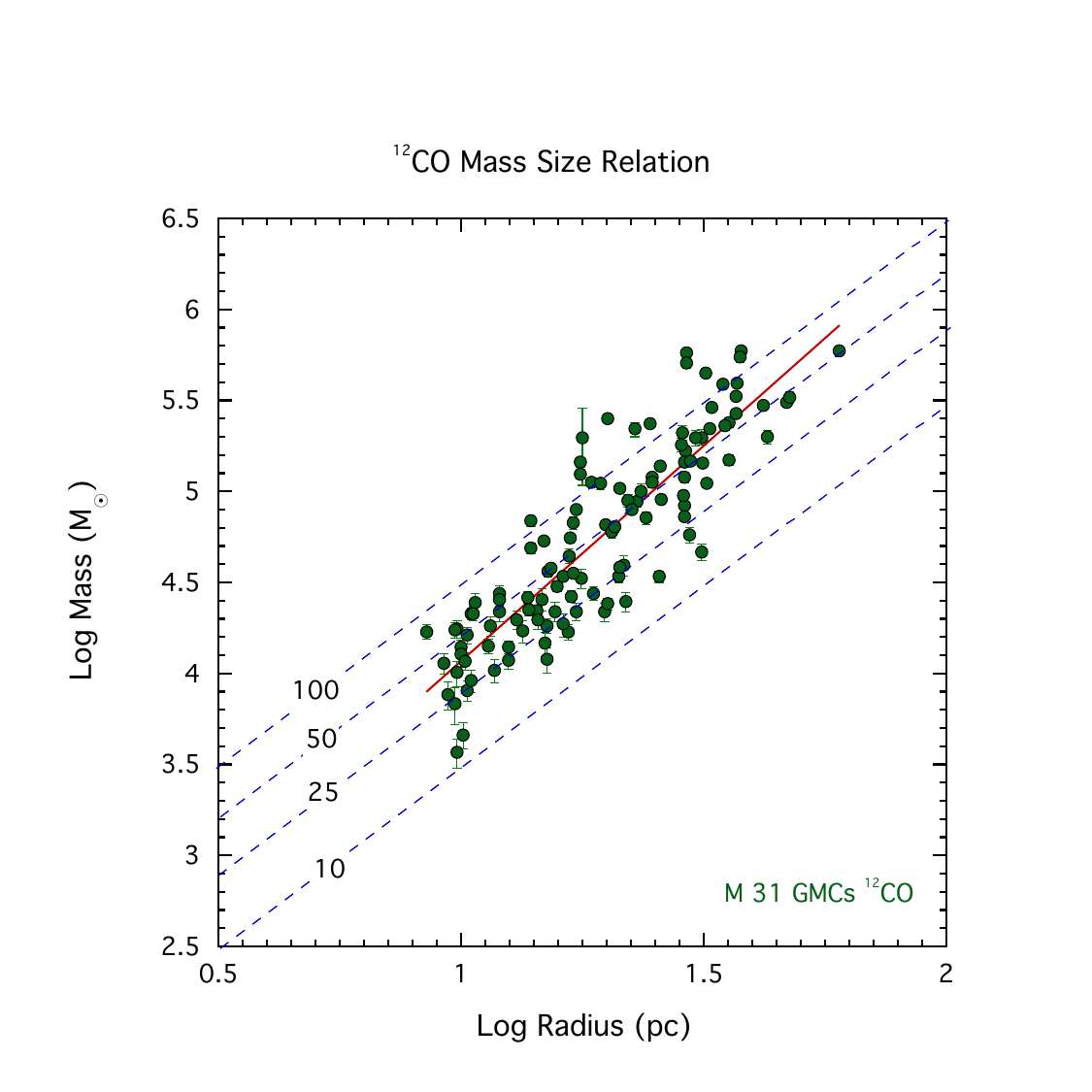}
\caption{Mass-Size relation for M31 GMCs derived from $^{12}$CO observations with the SMA of 117 GMCs with simple line profiles.  The dashed lines correspond to the loci of constant cloud surface densities (i.e., $\Sigma_{GMC}$ $=$ 10, 25, 50 \& 100 M$_\odot$pc$^{-2}$, respectively).  Formal error bars (also plotted ) are typically comparable or smaller in size than the plotted points. The red solid line is a simple least squares fit to the data, i.e., ${\rm log}(M_{GMC}) = 1.70 + (2.37\pm 0.12) {\rm log}(R)$.  \label{MvsR12all}  }
\end{figure}

Figure \ref{MvsR12all} shows the mass-size relation for the 117 M31 GMCs in our clean sample.  There is a clear, well defined, mass-size scaling relation for the M31 cloud population. A simple least squares fit to the data yields ${\rm log}(M_{GMC}) = 1.70 + 2.37 {\rm log}(R)$ with a correlation coefficient $R_{corr} = 0.88$ and a standard error (dispersion of residuals) of the fit of 0.25~dex.  Although we should expect some positive covariance between these two quantities because the masses implicitly depend on cloud size (i.e., $M_{GMC} = \Sigma_{GMC} \times \pi {\rm R^2}$), the fact that the correlation coefficient is close to unity indicates a  clear and strong correlation between the two variables.

 \begin{figure}[t!]
\centering
\includegraphics[width=0.6\hsize]{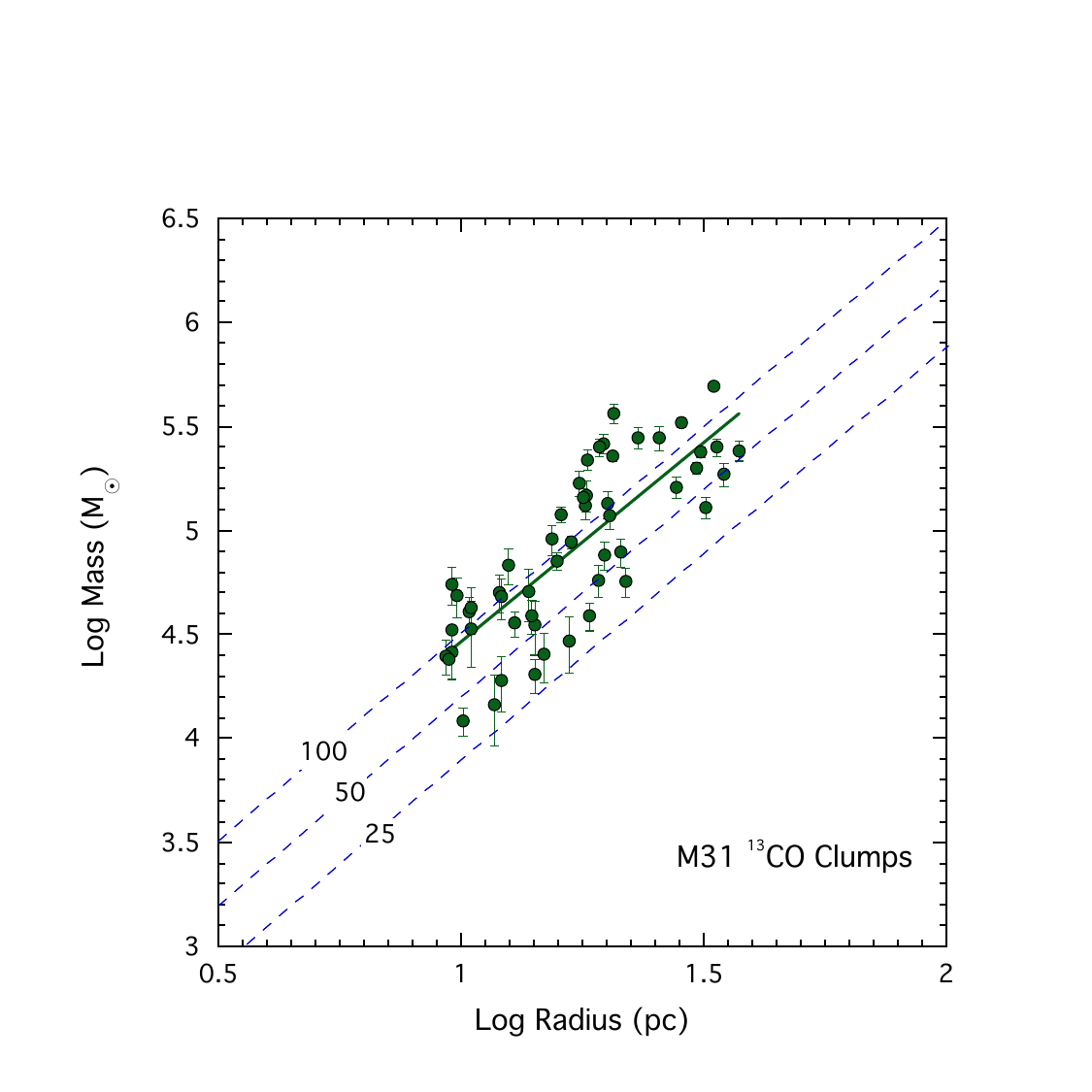}
\caption{Mass-Size relation for M31 $^{13}$CO emitting clumps. The dashed lines correspond to the loci of constant cloud surface densities (i.e., $\Sigma_{GMC}$ $=$ 25, 50 \& 100 M$_\odot$pc$^{-2}$, respectively). The green solid line is a simple least squares fit to the data, i.e. ${\rm log}(M_{\rm GMC}) = 2.55 + (1.92\pm 0.20) {\rm log}(R)$.   \label{13CO_MvsR_1G}  }
\end{figure}

The measured slope of the mass-size relation of  $2.37\pm 0.12$ is significantly different from 2. Although the GMC population of M31 clearly displays a strong mass-size relation, it is only approximately consistent with Larson's suggestion of a constant column density relation for GMCs. The scatter in the relation is appreciable and as can be seen in the figure the corresponding surface densities range over about an order of magnitude. However, the distribution of surface densities is not uniform over this range and we find $\sim 70$\% of the GMCs have surface densities between 20 - 70~\msun~pc$^{-2}$. Indeed, we directly measure a  median surface density of 48~\msun~pc$^{-2}$ for the entire clean sample while the mean surface density is found to be 57 $\pm$ 40 \msun~pc$^{-2}$.

We can also determine a mass-size relation using the \thco data. There are 53 \thco clumps associated with the restricted sample of 117 GMCs. Here again we find a relatively strong correlation between the two variables with a least squares linear fit yielding: ${\rm log}(M_{\rm GMC}) = 2.55 + 1.92 {\rm log}(R)$ with a correlation coefficient of 0.80 and a dispersion of the data about the fit of 0.25~dex. The slope of the \thco relation is 1.95 $\pm$ 0.20 and is, within the uncertainties,  Larson's  constant column density value.  The scaling or intercept value of the fit then corresponds to an average surface density of 100~\msun pc$^{-2}$,  about a factor 2 higher than the mean surface densities of the \twco gas. This is expected since the \thco emission arises from smaller spatial scales within the M31 GMCs than the \twco emission (see Viaene et al. 2021)
 and because of the density stratification of GMCs,  smaller spatial scales  translate to regions of higher surface and volume density. 
 Similar to $^{12}$CO, the \thco mass-size relation exhibits considerable scatter with corresponding surface densities ranging over about a factor of 8 in value, which is a somewhat lower dynamic range than that observed in the \twco relation. We find the mean surface density of the \thco sample to be $<\Sigma_{13CO}> = 103 \pm 56$~\msun pc$^{-2}$ with a median value of 92~\msun~pc$^{-2}$, consistent with the value derived immediately above from the fit intercept. 
 
 \subsubsection{Origin of the Observed Scatter and Implications for (internal) GMC Structure.}

It can be shown that  if stratified GMCs are characterized by power-law column density pdfs ($p(\Sigma) \sim \Sigma^{-n}$), like those in the Milky Way (e.g., Lombardi et al. 2015), their  mean column densities are given by 

\begin{equation}
< \Sigma_{GMC}> = {{n-1}\over{n-2}} \Sigma_0
\end{equation}

\noindent
 where $\Sigma_0$ is the outer boundary column density level and $n$ is the power-law index of the N-pdf (e.g., Lewis et al. 2022). That is, the mean column density derived for a GMC depends directly on the value of its boundary column density. Is this also true for the M31 GMC population? Can we learn something about the internal structure of the M31 GMCs from analysis of the mass-radius relation? 
 
\begin{figure}[t!]
\centering
\includegraphics[width=0.6\hsize]{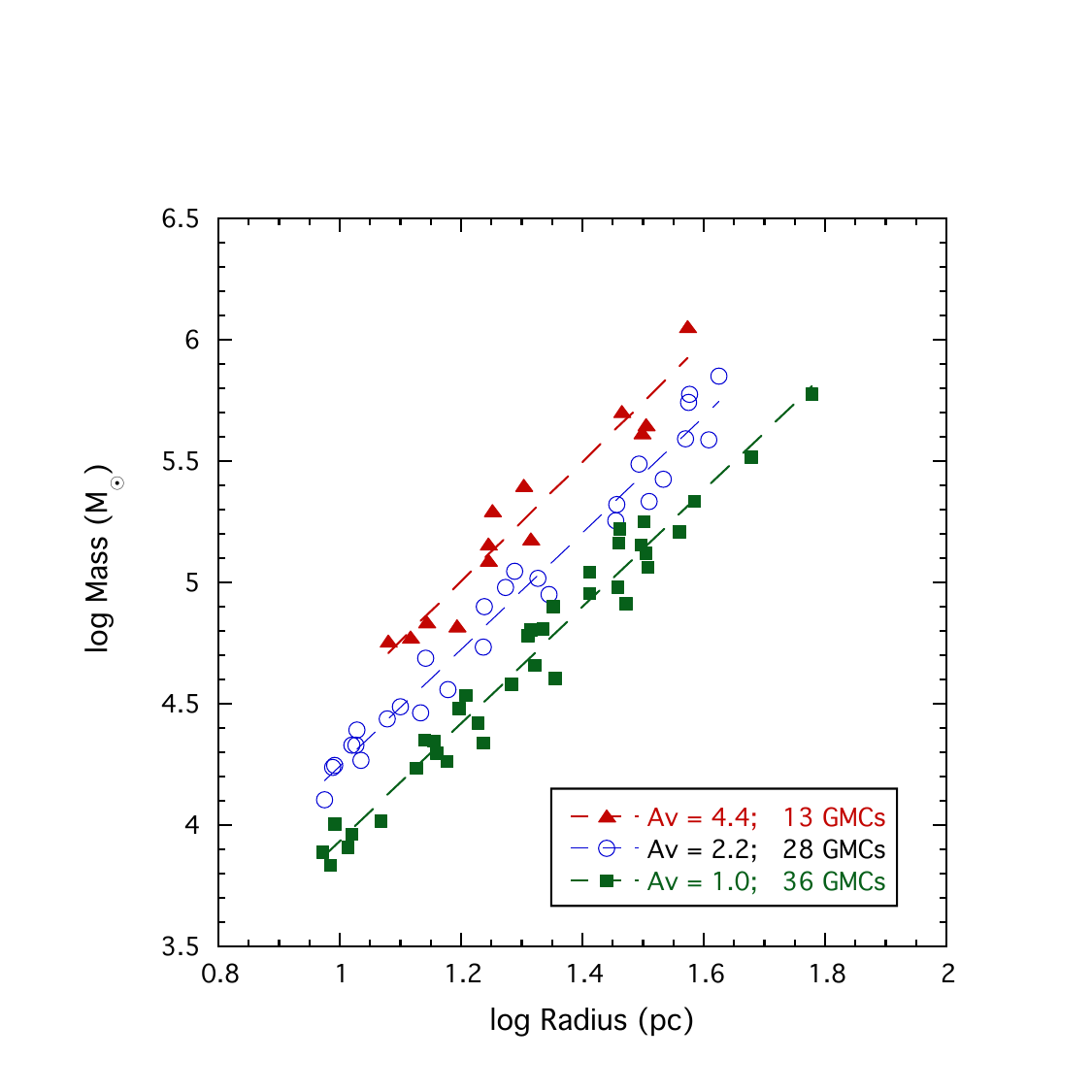}
\caption{Mass-Size relation for three sets of sources with three different 3 $\sigma$ boundary levels corresponding to extinctions of 1.0, 2.2 and 4.4 magnitudes (or $\Sigma_0$ of 22, 48 and 97 \msun pc$^{-2}$ ). For each boundary only sources whose boundaries were within 10\% of the given boundary were selected for the plot.    The dashed lines correspond to  simple least squares fits to the data (see text).  \label{MultiMvsRplot }}
\end{figure}

 To test this possibility we conducted an experiment to measure the mass-size relation for the M31 GMCs using groups of clouds for which the outer boundaries used to define them systematically differed.
We note here that although in our SMA survey the integration times were set to recover a fixed 3 $\sigma$ rms noise level ($\approx$ 0.2 mJy) in all the  wideband continuum images, 
variations in weather, bandwidth and the number of antennas operating at any one time produced some variations in the achieved target noise levels. 
As a result, the outermost cloud boundaries used to define the \twco clouds were not always exactly the same.  
Taking advantage of this circumstance we more closely examined the GMCs from our full 162 cloud sample. We selected three sub populations of GMCs taken from three sets of \twco images each with differing but internally precise values of image noise. The corresponding 3$\sigma$ boundary levels for these three GMC groups were, to a precision of 10\%,   $\Sigma_0$ = 22 $\pm$ 2,2, 48 $\pm$ 4.8 and 94 $\pm$ 9.4 M$_\odot$pc$^{-2}$ (equivalent to  A$_0$ =1.0 $\pm$ 0.1, 2.2 $\pm$ 0.22 and 4.4 $\pm$ 0.45 visual magnitudes). 
We plot in figure \ref{MultiMvsRplot }  the mass-size relations for these three groups. Linear fits to the individual relations are also plotted. The fitting parameters are listed in Table 3.

\begin{deluxetable*}{cccccc}
\tablenum{3}
\tablecaption{Fitting Parameters\label{tab:MultiFitparameters}}
\tablewidth{0pt}
\tablehead{
\colhead{$\Sigma_0$ (A$_V$)}&\colhead{Slope}&Dispersion (dex)&\colhead{R }&\colhead{$<\Sigma_{GMC}>$}&\colhead{${<\Sigma_{GMC}>\over \Sigma_{0}}$ }}

\startdata
22 (1.0)&2.42 $\pm$ 0.08& 0.09 &0.96&38&1.7 \\
48 (2.2)&2.40 $\pm$ 0.09& 0.09 &0.99&75&1.5 \\
97 (4.4)&2.51 $\pm$ 0.18&0.09&0.98&149&1.5 \\												
\enddata
\tablecomments{  Mass surface densities ($\Sigma$) expressed in units of \msun pc$^{-2}$} and extinctions (A$_V$) in units of magnitudes. Here  dispersion refers to the standard error of the fit and R is the correlation coefficient for the fit.
\end{deluxetable*}

The mass-size relations for the three sub populations of clouds each shows a strong correlation between the two variables with a correlation coefficient very close to 1.   
 The average mass surface densities of these populations are found to also scale linearly with the adopted boundary value, such that $\Sigma_{\rm GMC} = 1.6 \Sigma_0$. This indicates that the underlying structure of  the GMCs in M31 may be indeed characterized  by power-law pdfs or some other very steeply falling power-law--like function \citep[see][]{2012MNRAS.427.2562B}. If we assume that the underlying pdf for these clouds is a power law, we can use equation 1 to derive the index of the postulated power law and we find $n = 3.7$ for the M31 GMC population. This is precisely the value ($n = 3.66$) inferred for local Milky Way clouds from extinction measurements (Lewis et al.  2022).  However, we caution that this precise coincidence should not be taken too literally as Lewis et al. (2020) also showed that the N-pdfs of CO emission do not as accurately trace the underlying  structure of GMCs as the extinction pdfs. Nonetheless, these results do imply that the internal structures of GMCs in both galaxies must be  reasonably similar to each other.

 In addition our experiment finds that the scatter in each of the three relations is  nearly a factor of three lower than that of  the combined sample of 117 clouds (i.e., 0.09 vs 0.25).  This indicates that the bulk of the scatter observed in the \twco data in Figure \ref{MvsR12all} is a result of the variations in the adopted boundary levels. This is clearly evident in visual comparison of  Figures \ref{MvsR12all} and \ref{MultiMvsRplot }. 
At the same time the values of the slopes of the three relations are essentially identical to that derived for the full sample. Consequently,  the derived slope of the \twco mass-size relation of the M31 GMCs is quite robust. The magnitude of the residual scatter (0.09 dex) in each of the three relations exceeds  that of the formal errors in the data by a factor of about three, indicating that the scatter may contain additional information beyond the formal noise and may be dominated by sources of systematic uncertainty beyond variations in adopted cloud boundaries. Such factors as random spatial variations in molecular abundances or dust temperature across the galaxy could contribute to this level of scatter.  

\begin{figure}[t!]
\centering
\includegraphics[width=0.6\hsize]{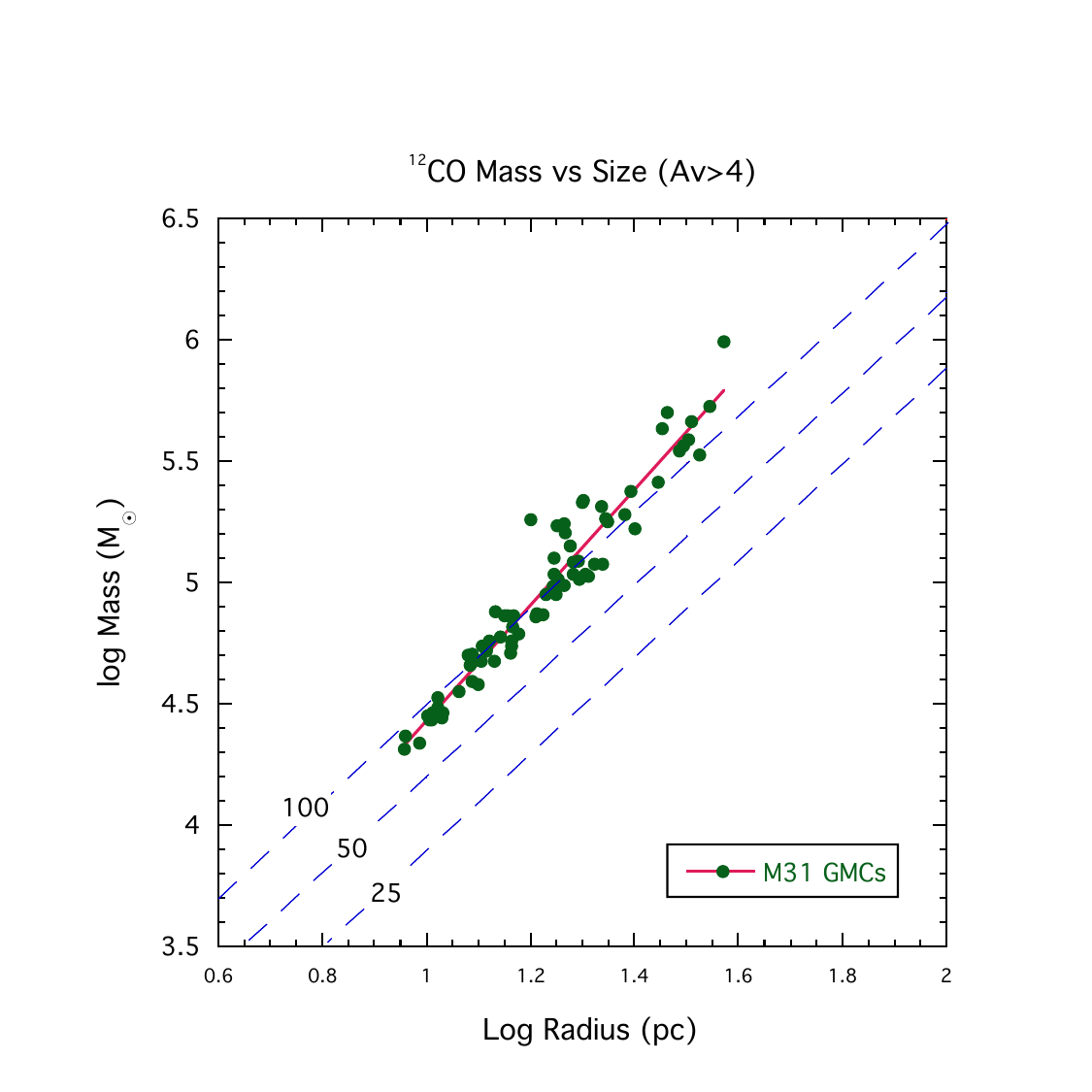}
\caption{Mass-Size relation for M31 GMCs derived from $^{12}$CO observations for clouds whose whose masses are calculated above a common fixed boundary column density that is equivalent to $\av = 4$ magnitudes. A solid red trace shows a simple least squares fit to the data that yields ${\rm log}(M_{\rm GMC}) = 2.37 + (2.38\pm 0.07) {\rm log}(R)$. Similar to  Figure \ref{MvsR12all}  error bars are typically comparable to or smaller in size than the plotted points and for clarity are not plotted here. \label{MvsR12AV4}  }
\end{figure}

As a cautionary note we point out that one might expect that some of the trends above could be introduced artificially as a result of using fixed surface density boundaries tied to the sensitivity limits of the observations. This is especially true for any faint clouds that would be marginally detected. Using a 3 $\sigma$ threshold as done here should help mitigate such concerns. Additionally we point out that such an artificially introduced trend among a set of noise ($\leq 1 \sigma$) peaks masquerading as marginally detected sources would produce a constant density mass-size relation with a slope of 2 and for which  $<\Sigma> = \Sigma_0$. This is not the case here. But as an additional test we determined the mass-size relation for the restricted sample using a fixed surface density boundary of $\av =$ 4 magnitudes. 
The result is shown in figure \ref{MvsR12AV4}. We adopted this boundary because it was 5-12 $\sigma$ above the noise level for the bulk  (70\%) of clouds in the sample. However only a subset of (76/117) of the clouds in the whole sample had peak surface densities that exceed this adopted boundary level and those are the GMCs that appear in the plot. The fitted relation for these sources is given by ${\rm log}(M_{\rm GMC}) = 2.37 + 2.38 {\rm log}(R)$ with a correlation coefficient $R_{corr}= 0.97$ and a dispersion of the fit residuals of 0.10~dex.  We find  $<\Sigma_{\rm GMC}>= 1.3 \Sigma_0$. These parameters are very similar to those in Table 3, consistent with the suggestion that the M31 GMCs are stratified with power-law or similarly structured, steeply falling,  N-pdfs.

\subsubsection{Comparison with Milky Way GMCs}

It is of interest to compare these results to the local Milky Way where molecular clouds can be studied in exquisite detail in both dust  and gas. Recently,  Lewis et al. (2022) performed a uniform and systematic study of both dust emission/extinction and \twco$(J=1-0)$  emission toward a dozen nearby GMCs. They found  tight power-law scaling relations between the masses and sizes of these GMCs. Using masses and sizes derived from \twco observations they derived a power law index of 2.22 $\pm$ 0.12 for the mass-size relation  close to what we found for M31. Moreover, the average surface density of the local clouds was found to be $<\Sigma_{\rm GMC}>\sim 38 \pm 8$ \msun~pc$^{-2}$ when calculated above an outer boundary of  $A_V\approx 1.0$ magnitude.  This is similar to the value (41$\pm$ 5) determined from extinction observations of local clouds (Lada et al. 2010) and essentially the same value found for the M31 GMCs in cases where a similarly deep outer boundary level can be used to define the cloud (see Table 3).  Moreover we note that the \twco mass-size relation for the local cloud sample derived by Lewis et al. had a dispersion of 0.10~dex, essentially identical to the dispersions listed in Table~3 for  M31 clouds  with precisely measured boundary levels and to that for the GMCs plotted in Figure \ref{MvsR12AV4}.  This suggests that for those GMC individual subsamples our SMA CO measurements have achieved the levels of depth and accuracy necessary to be competitive with Milky Way studies, that is, even deeper SMA observations would probably not substantially improve upon the results already achieved.

In another investigation
Lada \& Dame (2020) examined the nature of the mass-size relation for the entire Milky Way. From analysis of two independent GMC catalogs (Rice et al. 2016; Miville-Deschenes et al. 2017) each constructed from the complete \twco survey of the Galactic plane by Dame, Hartmann \& Thaddeus (2001). Lada \& Dame  found a mass-size relation for Milky Way GMCs with a slope of  approximately 2.1, again roughly consistent to that observed for the M31 GMCs. They also found that outside the molecular ring, which is located roughly between Galactic radii of $4 \leq {\rm R_{gal}} \leq 7$ kpc,  the GMCs could be well described by a nearly constant mass surface density of $\sim$ 35 \msun pc$^{-2}$, nearly identical to that found for local clouds and again quite close to what we find for M 31, especially considering  the likely differences in the values of the adopted outer boundary surface density levels between the two studies.  Within the molecular ring, Milky Way clouds have a radially dependent range of average cloud surface densities with values between 35 and $\sim$ 90 \msun pc$^{-2}$. However, severe overlap of clouds along the line-of-sight in the direction of the molecular ring produces an upward bias in most surface density measurements there making meaningful comparisons of Galactic Ring GMCs with M 31 and even local Milky Way clouds somewhat problematic. As pointed out earlier because of our vantage point, our M31 sample is free of such bias and our measurements are not subject to similar radially dependent experimental uncertainties. Nonetheless, it appears that the masses and sizes of the GMC population in M31 exhibit a very tight mass-size relation similar to that of GMCs in the local Galactic population and perhaps even more generally to those of GMCs across the Milky Way disk. Moreover, the extraordinary fact that the average surface densities of GMCs in the local region (i.e., D $\lesssim$ 1 kpc) of the Milky Way are even as similar as they are to those of GMCs across the disk of M31 is intriguing and suggests that the general physical environment  in which the two cloud populations find themselves is likely also very similar.

\subsection{Line width-Size (LWS) Scaling Relations}

The most well known and influential of Larson's relations is that between line width, or equivalently velocity dispersion ($\sigma$), and size, namely, 

$$ \sigma \sim R^p$$

\noindent
Larson found $p$ to be 0.38.  As Larson originally pointed out, the significance of this relation lies in its similarity to scaling relations for turbulent flows. In particular, the Kolmogoroff law, $\sigma \sim R^{0.33}$,  for incompressible, subsonic turbulence. Indeed, Larson's linewidth-size relation has often been cited as one of the key pieces of evidence indicating that molecular clouds are turbulent in nature (e.g., Larson 1981; V{\'a}zquez-Semadeni et al. 1997; Elmegreen \& Scalo 2004). However, subsequent studies of Galactic molecular clouds suggested $p$ $\approx$ 0.5 (e.g., Solomon et al. 1987, Garc{\'\i}a et al. 2014, Rice et al. 2016). This complicated the interpretation of the nature of the relation since the virial theorem predicts $\sigma \sim R^{0.5}$ for clouds with constant mean column density (e.g., Evans, 1999), apparently indicating that Larson's relations could be satisfied without the need for turbulent velocity fields. 

However calculations using compressible turbulence, a more appropriate form of turbulence for  GMCs, can produce $p$ = 0.5 (e.g., Passot et al. 2003). But these models are still not realistic because typically they did not include magnetic fields, gravity, or feedback, among the most critical pieces of physics for understanding the nature of  GMCs. The inclusion of gravity and magnetic fields in a simulation of magnetized, gravitationally stratified molecular clouds also has been able to produce $p$ = 0.5 (e.g., Kudoh \& Basu 2003) but without including feedback, a key piece of cloud physics. Due to the complex interplay of all these physical properties, it is not yet clear how much of a constraint the line width-size relation is for understanding the nature of turbulence in molecular clouds.

\begin{figure}[t!]
\centering
\includegraphics[width=0.6\hsize]{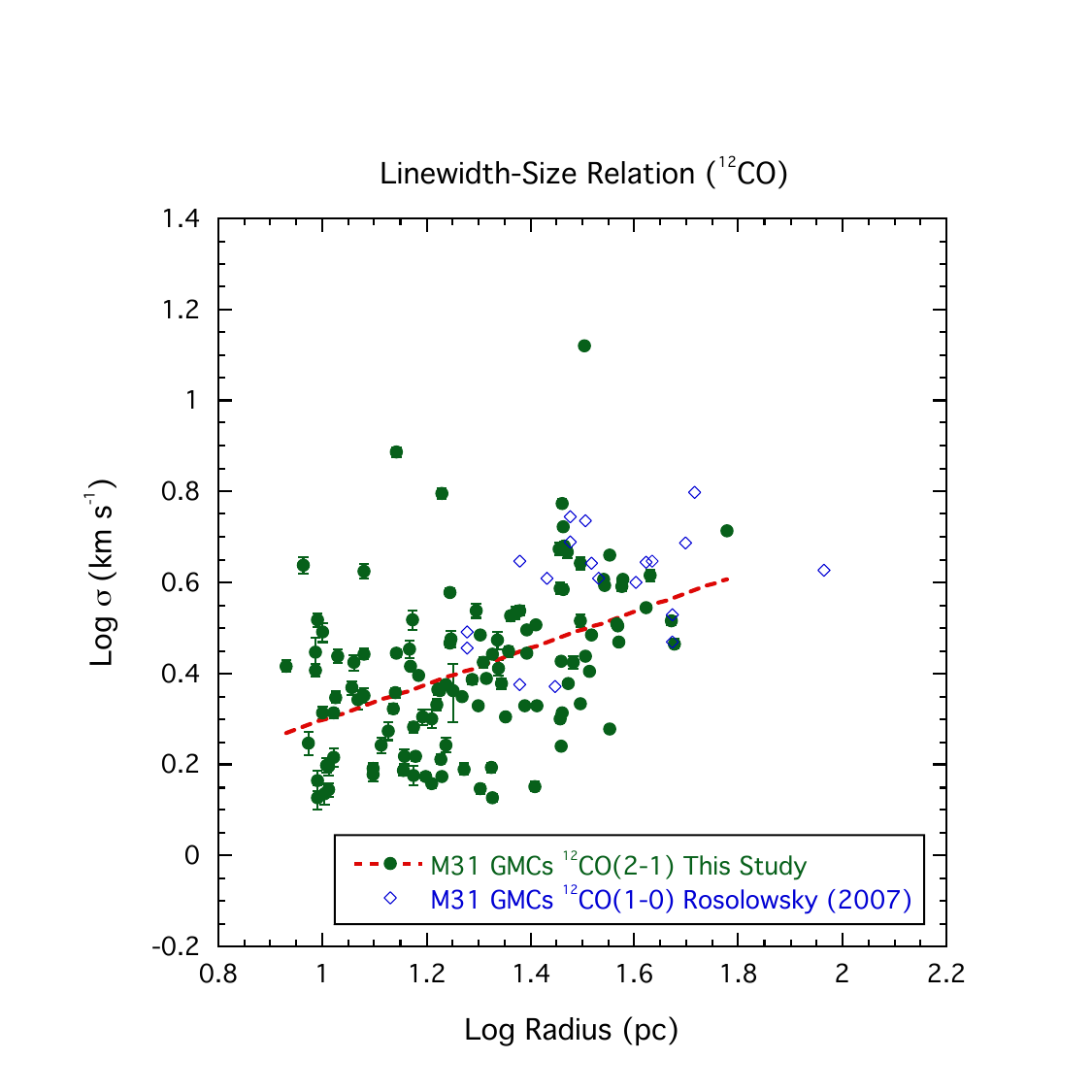}
\caption{Linewidth-Size relation for M31 GMCs derived from our $^{12}$CO(2-1) observations of clouds with simple gaussian line profiles (filled circles).  Error bars are displayed but are typically similar in size to the plotted circles. The dashed red line, a fit to the observations, is characterized by a slope = 0.40 $\pm$ 0.07. Also plotted  for comparison are $^{12}$CO(1-0) observations of M31 GMCs (open symbols) from the study of Rosolowsky (2007). See text for discussion. \label{LWS12CO1G} }
\end{figure}

In Figure \ref{LWS12CO1G} we plot the linewidth-size relation (LWS) for the M31 GMCs  in our restricted SMA sample. There appears to be a modest correlation between the velocity dispersion and size in the SMA sample, although the scatter is large.  A least-squares fit to these data is plotted on the figure as a short (red) dashed line and yields a slope of 0.40 $\pm$ 0.07 with a correlation coefficient of 0.44 and a dispersion of 0.16~dex around the fitted relation. The derived slope of the correlation compares well with that originally derived by Larson and is formally  lower than the value (0.5) quoted in most Milky Way studies. However, the dispersion in the fitted values and the low correlation coefficient  are indicative of a relatively weak correlation between these two variables. Indeed, as is the case here for M31, the LWS relation is often the least well correlated of the Larson scaling relations for cloud populations within individual galaxies as prior observations of  M33 (Gratier et al. 2012), M51 (Colombo et al. 2014) and NGC 300 (Faesi et al. 2018), and even the Milky Way  (e.g., Larson, 1981;  Heyer \& Dame 2015) have indicated.  This may not be so surprising since the linewidths intrinsically only weakly vary with size ($\sim$ r$^{0.4-0.5}$) in the first place. This correlation may not be particularly interesting or influential on scales of individual GMCs.

\begin{figure}[t!]
\centering
\includegraphics[width=0.6\hsize]{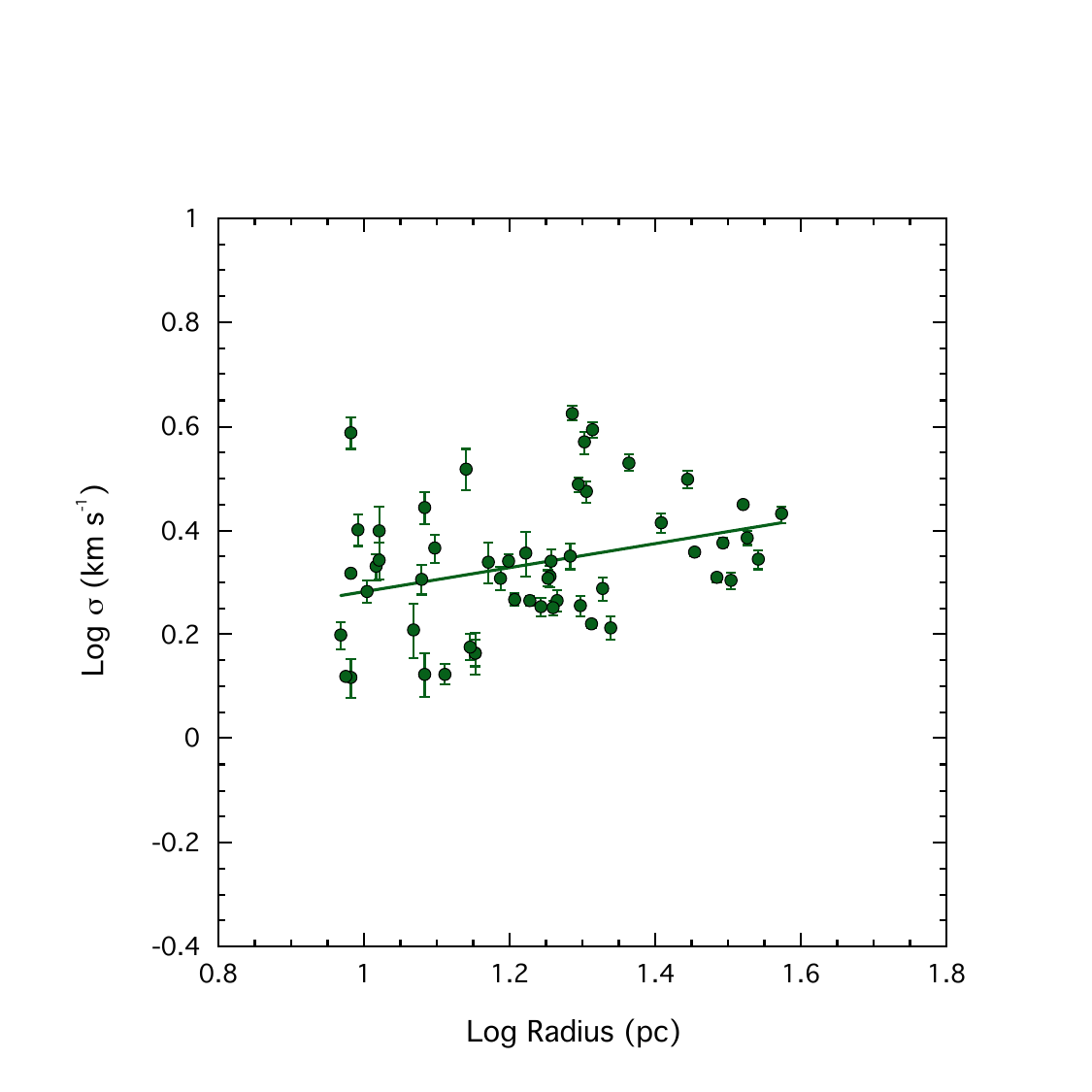}
\caption{Linewidth-Size relation for M31 GMCs derived from our $^{13}$CO(2-1) observations. The solid line is a fit to the observations and yields a slope of 0.23 $\pm$ 0.10 for the fitted relation.   \label{LWS13CO1G} }
\end{figure}

For comparison, with our SMA data we also plot interferometric \twco(1-0) data of M31 from Rosolowsky (2007) that were obtained with the Berkeley-Illinois-Maryland Array (BIMA) telescope. Despite experimental differences in beam sizes ($\sim 8''$ vs $\sim 4''$), velocity resolution (2.0 vs 1.3 km\,s$^{-1}$), cloud definition used and transition ($J=1-0$ vs $J=2-1$) observed, the BIMA and SMA distributions  agree quite well where they overlap on the LWS diagram. This is likely due to the fact we are dealing with the large scale and integrated properties of the GMCs in M31 and intrinsically the LWS is not a particularly strong correlation.  Nonetheless, given the differences enumerated above, we hesitate to make a more quantitative comparison of these two data sets.

Since we also have extensive \thco observations  we can investigate the LWS relation for  the M31 GMCs in \thco emission. In Figure \ref{LWS13CO1G} we show the LWS relation for our cloud sample. In \thco emission the velocity dispersion is only very weakly correlated with size. A linear fit to the data is plotted and returns a slope of 0.23 $\pm$ 0.10 with a correlation coefficient of only 0.32 and a dispersion of 0.12 dex around the predicted relation. The correlation is  weaker than that seen in \twco emission. In particular, the slope of the relation is less than that (i.e., $\sim$ 0.5) reported for \twco in the Milky Way and other galaxies. Indeed, this data does not provide much support for the existence of a significant LWS relation in the \thco emission. In this context it is of some interest to compare our result to \thco observations of GMCs in the Milky Way. Perhaps the most extensive \thco data available for Milky Way GMCs is that obtained in the Galactic Ring Survey (Jackson et al. 2006) and analyzed and published by Roman-Duval et al. (2010). We constructed the linewidth-size relation for that data  and found very similar results to M31. Namely, a very weak correlation between the two variables with a least squares linear fit yielding a slope of 0.27 $\pm$ 0.02 with a correlation coefficient of 0.55 and a dispersion of the data about the fit of 0.19~dex. 

The origin of the differences between the LWS relations for \twco and \thco  in M 31 is not clear but perhaps our results suggest that the strength of the LWS relation may depend on spatial scale. At first glance this may not be surprising since we expect turbulence to decay as scale size in the cloud decreases so that on the scales of dense cores there is no linewidth-size relation  and very little turbulence (e.g., Lada et al. 2008; Kirk et al. 2017; Hacar et al. 2013, 2018). However, this may not be the full explanation for the possible  scale dependence of the LWS relation observed here because on the spatial scales of the \thco emission the velocity dispersions  we observe (i.e., 1-3 km\,s$^{-1}$) are still supersonic and dominated by non-thermal motions. Perhaps this indicates that gravitational contraction becomes an important dynamical influence on the measured velocity dispersions on the size and density scales traced by \thco.

\begin{figure}[t!]
\centering
\includegraphics[width=0.6\hsize]{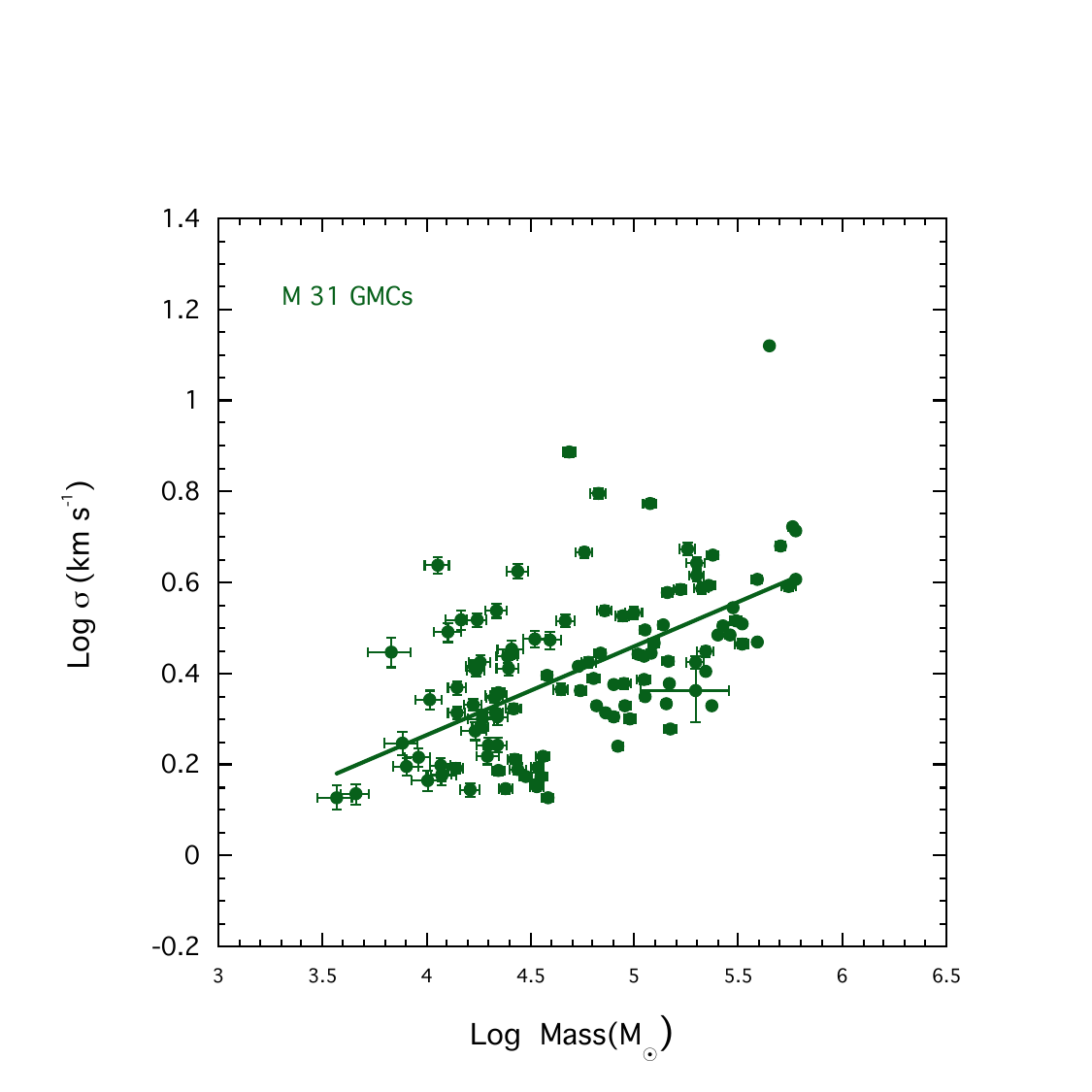}
\caption{Velocity dispersion-Mass relation for M31 GMCs derived from our $^{12}$CO(2-1) observations. The solid line is a fit to the observations. \label{sigmaMass1G} }
\end{figure}

\subsection{The  Virial vs. Luminous Mass Scaling Relation. Are M31 GMCs Gravitationally Bound?}

To investigate the physical influence of cloud masses (and gravity) on their dynamics Larson plotted the relation between cloud velocity dispersion and mass. He found that $\sigma \sim M^{0.2}$ with a dispersion of 0.12~dex about the relation. In Figure~\ref{sigmaMass1G} we show the velocity dispersion mass diagram for our restricted M31 sample. For M31 we find $\sigma \sim M^{0.20}$ with a dispersion about the fit of 0.14~dex and a correlation coefficient of 0.59. Both the slope and rms deviation of points around the fit are essentially identical to Larson's values for the Milky Way. Larson surmised that this relation might indicate that clouds were in a dynamical state close to virial equilibrium. Consider for example, the virial relation, $\sigma^2 \sim GM/R$ and the mass-size relation, $M \sim R^2$,   if $R \sim M^{0.5}$ then $\sigma \sim M^{0.25}$, close to what Larson found. 

Larson then plotted the virial parameter against cloud size and found a very weak relation indicating that the Galactic clouds were likely in a state close to being virialized and thus were most likely bound. Supporting evidence for this interpretation for gravitationally bound or even virialized clouds has been claimed by many subsequent studies (e.g., Solomon et al. 1987; Blitz \& Williams 1999; Heyer et al. 2009; Heyer \& Dame 2015; Garcia et al. 2014). However, others have argued that clouds may not be bound gravitationally. For example, Maloney (1988, 1990) suggested that clouds were bound by external pressure of the ISM and not necessarily gravitationally bound or viralized with only the gravitational potential and kinetic energy. Most recently work by Evans et al. (2021) has provided what appears to be direct evidence suggesting that most (70-80\%) Milky Way clouds measured in \twco are unbound.

\begin{figure}[t!]
\centering
\vskip -0.9in
\includegraphics[width=1.0\hsize]{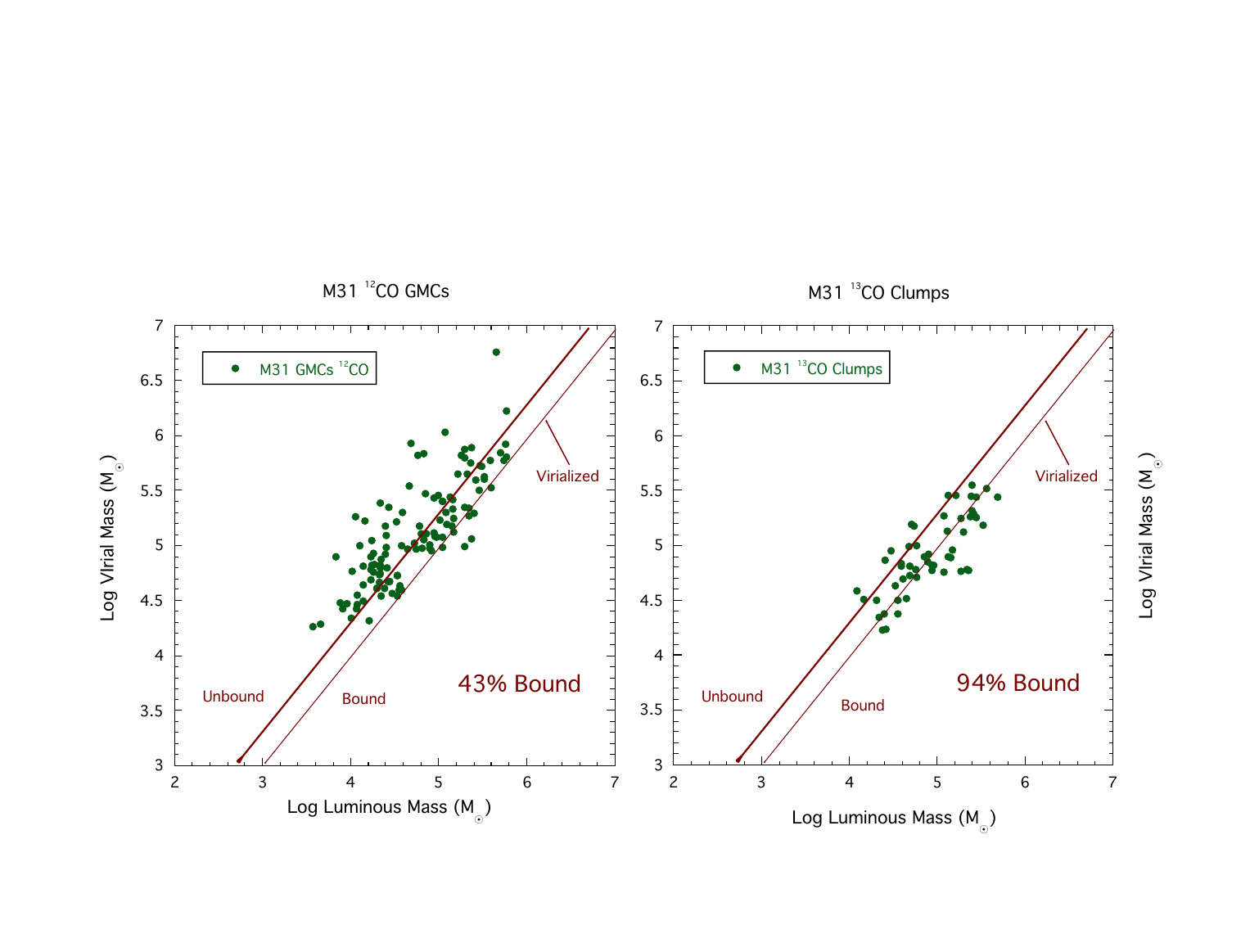}
\caption{Virial masses plotted against a) the \twco and b) \thco luminous masses for the M 31 GMCs with simple line shapes. The virial condition is represented as a solid line and the border separating gravitational bound and unbound GMCs as a thick solid line. The diagram indicates that only 43\% of the \twco luminous GMCs in our M31 sample are likely bound.  In contrast 94\% of the \thco luminous clumps appear bound with most scattered relatively tightly around the virial relation. For this figure the luminous cloud masses were derived using the corresponding dust calibrated CO conversion factors, $\alpha$(\twco) and $\alpha$(\thco) derived from earlier observations of the M31 clouds (see text).  \label{MvirVsMtot1G}}

\end{figure}

Empirically the virial nature of GMCs has often been inferred from or demonstrated by either 
 a one-to-one scaling relation between the virial and luminous masses of a cloud population, that is, $M_{virial} = M_{lum}^{1.0}$ or the finding that the virial parameter, $\alpha_{vir},  {\rm is} \approx 1$. 
 In Figure~\ref{MvirVsMtot1G} we plot the relations between the virial and the $^{12}$CO and \thco luminous masses of the 117 GMCs and 53 \thco clumps in our M31 restricted sample. We determined the virial mass of each GMC using the equation
 $M_{vir} = 1040 \sigma^2 R$, which assumes a stratified cloud with an internal density gradient given by  $\rho(r) \sim r^{-1}$  (Solomon et al. 1987). The corresponding virial parameter is then $\alpha_{vir} = {4.5\sigma^2R\over{GM}}$.\footnote{This differs from the standard expression, $\alpha_{vir} = {5\sigma^2R\over{GM}}$, which assumes a constant density cloud.}  The plot shows that not only are most of the GMCs located above the virial relation but most of these clouds lie in the unbound region of the diagram. Indeed, we find that only 43\% of the M31 GMCs  are likely gravitationally bound. We note that  whether or not a cloud is bound is mass dependent.  Of the 41 GMCs with luminous masses exceeding $10^5$ \msun, 71\% (29/41) are bound while only 27\% (21/76) of the remaining lower mass GMCs are bound. In contrast,  the \thco clumps mostly cluster around the virial relation with 94\% of the clumps being bound.
 
To explore the sensitivity of this result to the adopted parameters of the dust derived CO mass calibration ($\alpha_{CO}{\rm(T_D)}$) and stratification of the GMC (i.e., $\rho(r) \sim r^{-\beta}$)  we have also calculated the bound fractions assuming different mass calibrations corresponding to plausible ranges of dust temperature (15 K -- 25K) and different possible values of $\beta$. The results are listed in Table 4. It is clear from inspection of Table 4 that some caution needs to be applied when evaluating the boundedness of a cloud or cloud population based on standard virial analysis. For a fixed $\beta$ the derived bound fraction decreases with increasing dust temperature whilst for a fixed dust temperature calibration, the bound fraction tends to increase with $\beta$. For the M31 population studied here, {\it Herschel} observations find dust temperatures on GMA scales to be typically  $\sim$ 18 K, with $\sim$ 75\% of the GMAs characterized by temperatures in the range 15-20 K (Kirk et al. 2015). 
As discussed earlier the GMCs in M31 are likely stratified similar to clouds in the Milky Way with the most plausible value of $\beta$ likely being 1 (Solomon et al. 1987) which we adopt here.

 The results, that most GMCs in M31 appear gravitationally unbound  when traced in \twco emission, are in contrast to long accepted findings for the Milky Way mentioned earlier (e.g., Larson 1981; Solomon et al. 1987;  Blitz \& WIlliams 1999, Heyer et al. 2009, etc.). However,  our results are qualitatively similar to the recent findings of Evans et al. (2021) for the Milky Way. Evans et al. examined six \twco catalogs of Milky Way clouds derived from four separate surveys and estimated the bound gas fractions for each catalog, assuming $\alpha_{vir}$ for a spherical cloud (i.e., $\beta = 0$). They found the average bound fraction of molecular clouds in these catalogs to be $<$f$_{bnd}$(N)$>$  = 0.22 $\pm$ 0.16. Similarly they examined five \thco cloud catalogs but found $<$f$_{bnd}$(N)$>$  = 0.60  $\pm$ 0.17 indicating that most of the \thco clouds in the Milky Way were bound. These values compare to our M31 measurements of 0.43 and 0.94, respectively (using our nominal calibrations).

\startlongtable
\begin{deluxetable*}{cccc}
\tablenum{4}
\tablecaption{GMC Bound Fractions\label{tab:boundfrac}}
\tablewidth{0pt}
\tablehead{
\colhead{T$_D$ (K)}&\colhead{$\beta =$ 0}&\colhead{$\beta =$ 1 }&\colhead{$\beta =$ 2}}
\startdata
15&54\%&58\%&78\% \\
18&42\%&{\bf 43\%}&65\% \\
20&31\%&37\%&59\% \\	
25&14\%&24\%&43\% \\											
\enddata
\tablecomments{Bound fractions for GMCs with $\rho$(r) $=$ $\rho_0$ r$^{-\beta}$ and GMC masses calculated for different dust temperatures, T$_D$. The bold faced entry corresponds to the parameters used for this study (see text).}
\end{deluxetable*}

Also similar to our findings, Evans et al.  noted that the fraction of bound clouds increased with cloud mass implying that the fraction of bound clouds by mass would be larger than the fraction of bound clouds by number, that is, f$_{bnd}$(M) $>$ f$_{bnd}$(N). This is certainly the case for M31 where we find f$_{bnd}$(M) = 0.66 for the \twco emitting GMCs. This compares to the average for the 6 Milky Way ($^{12}$CO) catalogs Evans et al. investigated, $<$f$_{bnd}$(M)$>$ = 0.52 $\pm$ 0.26.  
Although the majority (57\%) of the GMCs in our M31 sample are not gravitationally bound, most of the mass of the molecular gas contained in the GMC population is still  within bound structures and substructures, qualitatively similar to the Milky Way. This is not surprising since the GMC mass functions of both the Milky Way and our M31 sample both fall with increasing GMC mass in a manner requiring most of the mass to be tied up in the most massive objects. 

In terms of their stability the molecular cloud population in M31 seems similar to that of the Milky Way and the Larson paradigm that GMCs are gravitationally bound objects may not hold up in either galaxy. Although, as we have already mentioned and will discuss in more detail below, there is a particular population of GMCs in M31 which are well described by Larson's paradigm.

\begin{figure}[t!]
\centering
\includegraphics[width=0.6\hsize]{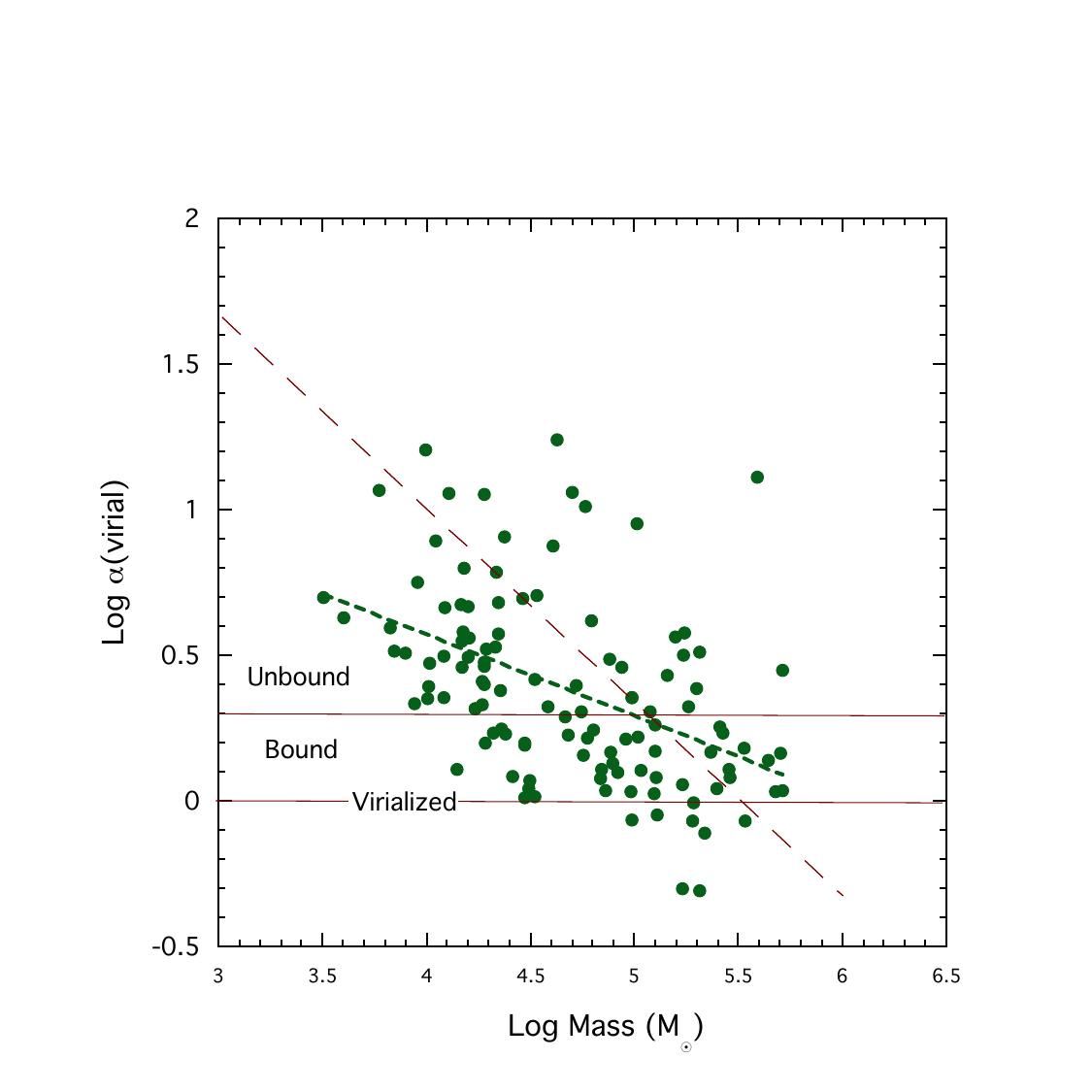}
\caption{Virial parameter plotted against GMC Mass for the GMCs detected in \twco with simple profiles. The short dashed line is a linear fit to the data. The long dashed line is a line with the slope of the virial relation expected for pressure confined clouds. (see text)  \label{alpha_vs_M} }
\end{figure}

\subsection{Are GMCs Pressure Confined?}

Although the GMCs in M31 mostly appear unbound, a long debated possibility remains that they could still be in a state of dynamical equilibrium by being confined by the external pressure from such pressure sources as the intercloud medium (e.g., Maloney 1990, Bertoldi \& McKee 1992),  the weight of the stars in a galaxy disk (Elmegreen 1989; Blitz \& Rosolowsky 2004), a combination of such effects (Sun et al. 2020).   Figure \ref{alpha_vs_M} displays the relation between the virial parameter and the mass of a GMC. The figure confirms our earlier analysis that a majority of GMCs are not gravitationally bound and only a few of the more massive objects are near virial equilibrium.  In addition $\alpha$(vir) appears modestly correlated with GMC mass. A linear fit to the data finds a slope of 0.28 $\pm$ 0.03 with a correlation coefficient of 0.46 and a dispersion of 0.28 dex. Also plotted is a solid trace with the slope expected  for pressure-confined, uniform density clouds in virial equilibrium including surface pressure terms (Bertoldi \& McKee 1992).  The data only roughly approximate the expectation of pressure-confined GMCs in virial equilibrium. Nonetheless given the large scatter and uncertainties and the fact that real GMCs are not of uniform volume density as assumed in the theory, we cannot with the present data rule out the possibility that the M31 GMCs are pressure-confined entities.

\subsection{Two Classes of GMCs: Further Insights into the Larson Relations}

From our initial segmentation analysis we identified two classes of GMCs, those with strong \thco emission and those with no or weak \thco emission. We found 44 GMCs in our restricted sample to be strongly  \thco emitting and 73 to be weakly \thco emitting. We refer to the \thco bright GMCs as dense GMCs and those with weak or no \thco emission as diffuse GMCs, similar to the nomenclature introduced by Roman-Duval et al. (2016). However, in our case the diffuse \twco clouds are discrete, identifiable objects. Spatially diffuse \twco emission may also exist in our fields but is likely filtered out by the interferometric response. We also note that some of the GMCs that are identified as diffuse in a moment~0  image do show weak but detectable \thco emission in the deeper, spatially integrated spectra while many more show no detectable \thco in the spectra. 

\begin{figure}[t!]
\centering
\includegraphics[width=0.9\hsize]{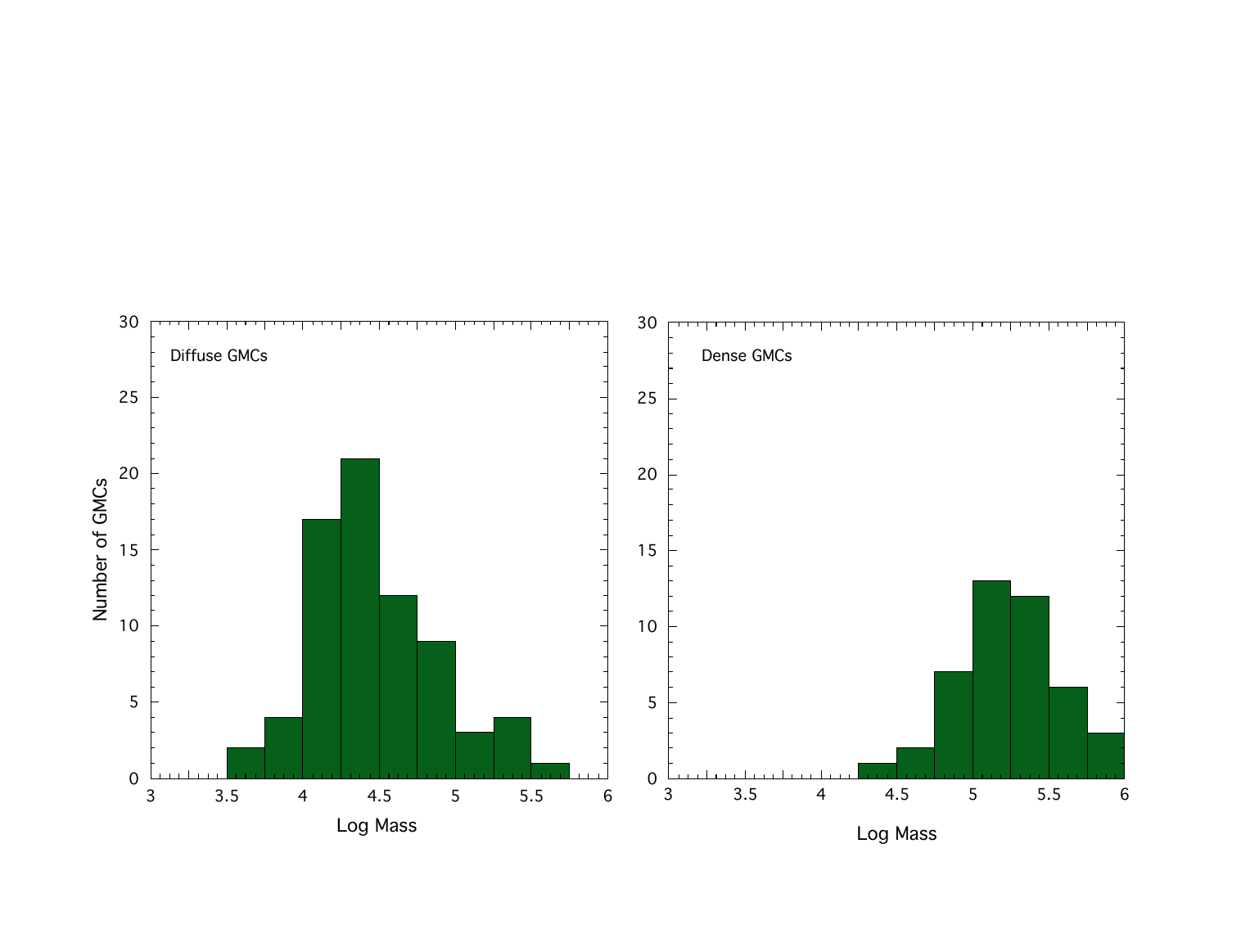}
\caption{Frequency distributions of masses for the diffuse  (left) and dense (right) GMC populations. The strong \thco emitting, dense GMCs constitute the bulk of the GMCs with masses in excess of 10$^5$ \msun. (see text). 
 \label{Masses_dense_diffuse}}
 \end{figure}

In Figure \ref{Masses_dense_diffuse} we display the mass distributions of the two GMC populations. The two populations are clearly different in their mass distributions. The strong \thco emitting, dense GMCs represent a population of more massive objects than the weak \thco emitting, diffuse GMCs. Indeed, 80\% of the GMCs with masses in excess of 10$^5$ \msun\  are dense GMCs. The average masses of the dense and diffuse GMC populations are $<{\rm M_{GMC}}>_{\rm Dense} =$ 2.2 $\times$ 10$^5$ \msun\  and $<{\rm M_{GMC}}>_{\rm Diffuse} =$ 4.8 $\times$ 10$^4$ \msun, respectively. 

\begin{figure}[t!]
\centering
\includegraphics[width=0.6\hsize]{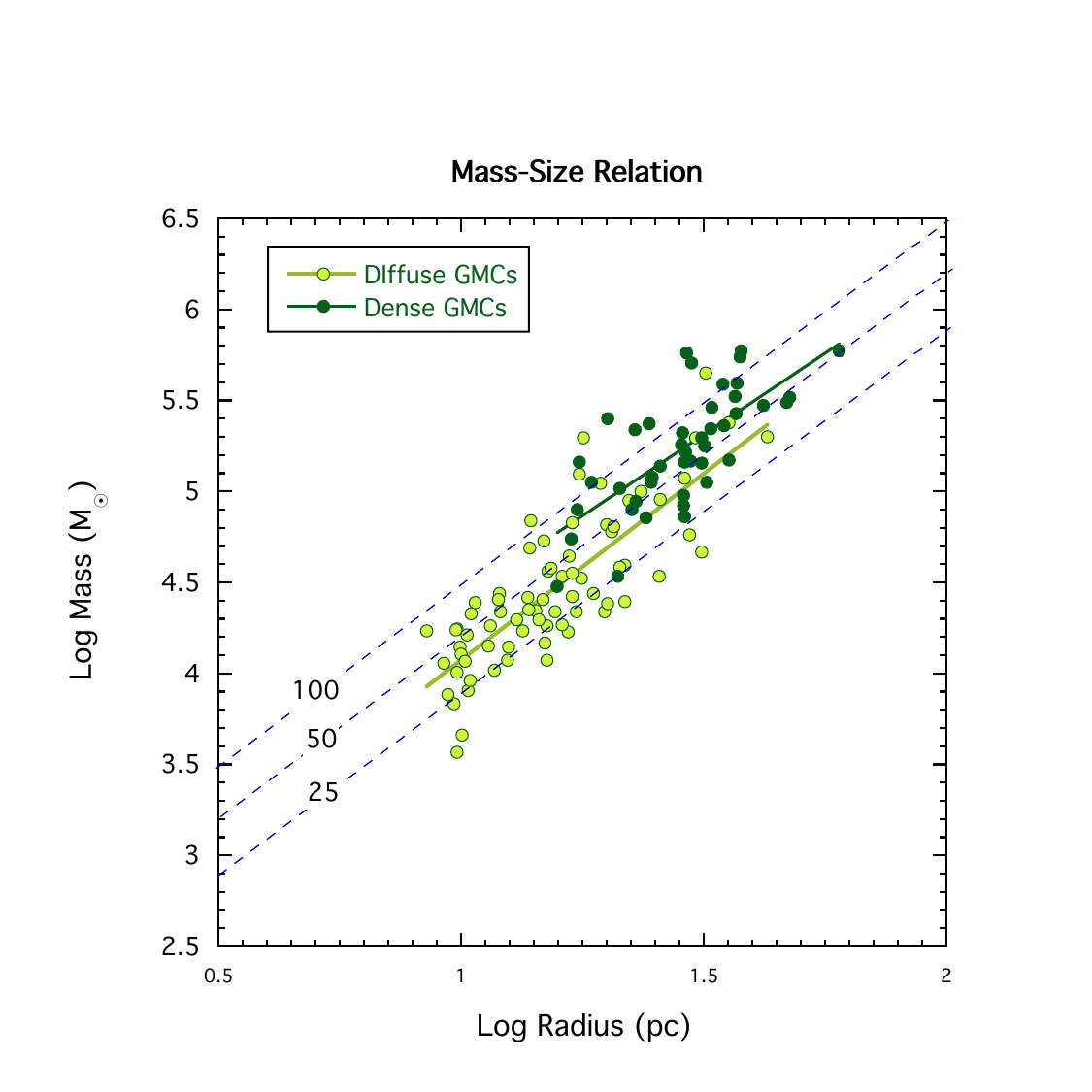}
\caption{Mass-size diagram illustrating the difference between the dense and diffuse GMC populations in M31 in this parameter space. The dense, $^{13}$CO bright, GMCs are represented by the dark green filled circles while the diffuse, $^{13}$CO faint,  GMCs are indicated by light green filled circles. Solid lines are the linear least-square fits to the corresponding dense and diffuse populations. (see text). 
 \label{MvsR12+13}}
 \end{figure}

In terms of the Larson relations, these two populations display some interesting similarities and differences.   As shown in Figure~\ref{MvsR12+13} we found that both classes separately produced mass-size relations with similar slopes,  specifically, 1.79 $\pm$ 0.27  and 2.05 $\pm$ 0.18 for the dense and diffuse GMCs respectively.  However, the two mass-size relations are offset from each other in the mass-size plane. This corresponds to the two populations having differing mean surface densities; specifically,  $<\Sigma_{\rm GMC}> = 77 \pm 7$ and $46 \pm 4 M_\odot$\,pc$^{-2}$, for the dense and diffuse populations, respectively\footnote{Note here the quoted errors are the standard errors in the means}. Moreover, individually the slopes of the two populations are  decidedly shallower than the value ($\sim$ 2.4) characterizing the combined population (figure \ref{MvsR12all}) and, within the errors, both consistent with the Larson constant column density value of 2.0.

Within the uncertainties the slopes of the  linewidth-size relations for the dense and diffuse GMCs  (0.47 $\pm$ 0.14 vs 0.43 $\pm$  0.13) are the same and within the errors the same as that of the combined population. However, the observations of the diffuse GMCs do show more scatter and are even less correlated than those of the dense GMCs. Nonetheless, both populations have poorly correlated LWS relations, similar to the findings for the combined population as a whole.

\begin{figure}[t!]
\centering
\includegraphics[width=0.6\hsize]{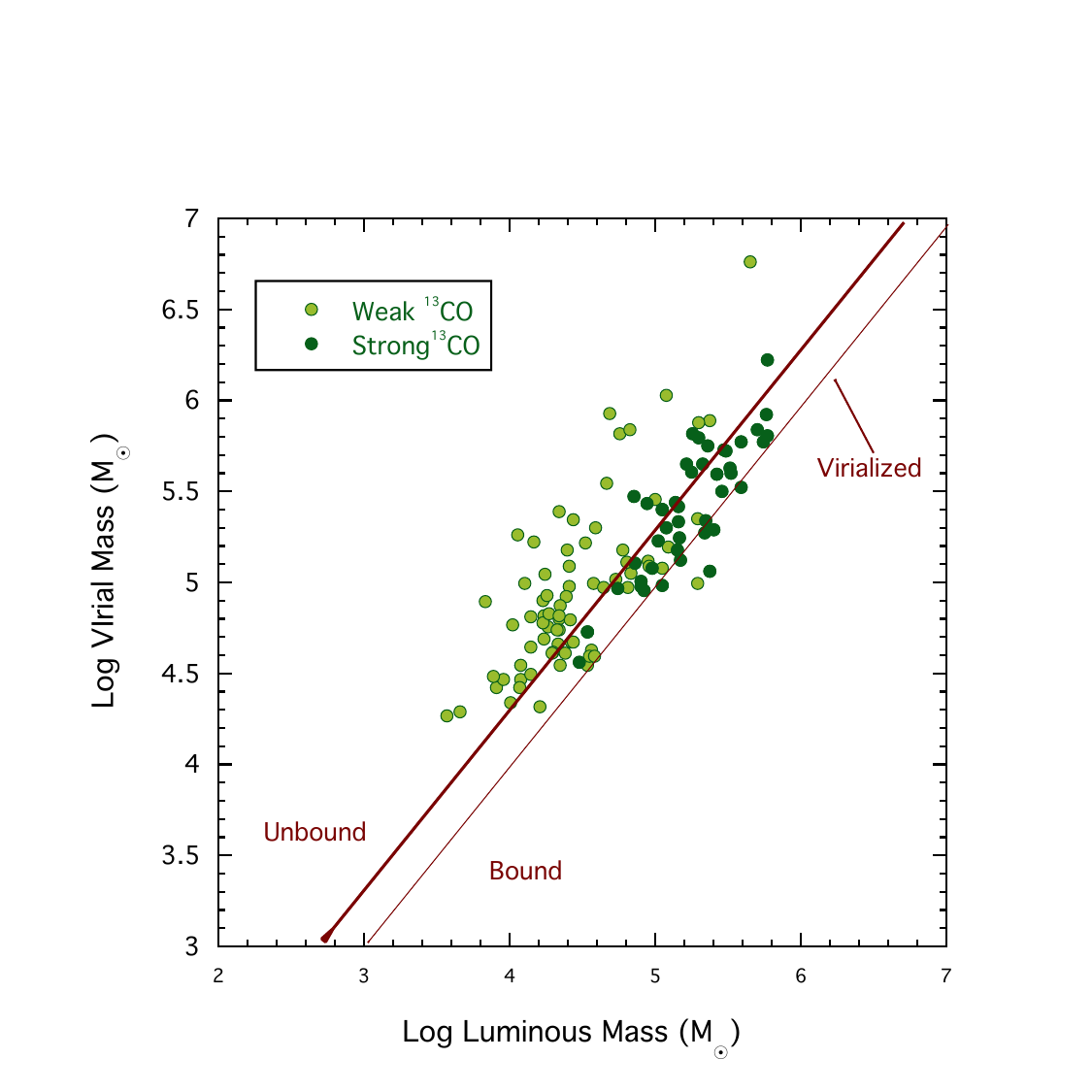}
\caption{Virial masses plotted against the \twco luminous masses for the M 31 GMCs with simple line shapes.  The dense, $^{13}$CO bright, GMCs are represented by the dark green filled circles while the diffuse, $^{13}$CO faint,  GMCs are indicated by light green filled circles. The figure shows the weak \thco emitting GMCs to be mostly unbound while the strong \thco GMCs are mostly bound (see text).
 \label{Mvir_vsMlum_12+13}}
 \end{figure}

Finally we find that the dense and diffuse clouds are characterized by significantly different degrees of gravitational boundedness. Figure~\ref{Mvir_vsMlum_12+13} shows the relative locations of the dense and diffuse GMCs on the virial vs luminous mass diagram. We find the fraction of dense GMCs that are bound to be 73\% while the bound fraction of diffuse GMCs is found to be 25\%. This is not surprising given our earlier findings on the bound fraction of  GMCs with masses in excess of 10$^5$ \msun\  and the fact that the strong \thco emitting GMCs consist of the most massive GMCs in our sample. 
Because the diffuse GMCs are mostly unbound they at most follow only two of Larson's relations. The dense GMCs appear to conform to all the Larson relations. Being gravitationally confined the dense GMCs are also likely to be the most active star forming clouds in our sample. Support for this surmise derives from the fact that star formation in galactic disks has a particular affinity for the higher density confines of GMCs (Lada et al. 2010, Heiderman et al 2020,  Nuemann et al. 2023). Recent inteferometric HCN observations of a subset of our M31 sample suggest that the dust cores detected by our SMA observations likely correspond to the densest regions of the GMCs in our survey (Forbrich et al 2023). Of the 40 dust cores we identified in the sample, 38 are found in the dense, \thco bright, GMC population. 

In terms of Larson's relations and star formation the diffuse GMCs may be the most interesting population. 
An intriguing aspect of the Larson relations has been the property that given gravitational bound clouds, either the mass-size relation or the linewidth-size relation follows directly from the other. This would mean that there is only a need to establish a physical basis for one of these two relations.  It has often been assumed that the linewidth-size relation is the physically motivated relation, given our understanding of the  turbulent nature of GMCs, an understanding derived in no small part from the existence of the  linewidth-size relation itself!  However, for a population of GMCs that are unbound, the linewidth-size and mass-size scaling relations are each required to have an independent physical explanation or basis. 
What are the underlying physical origins for these two relations in an unbound population of clouds? It may be that both these scaling relations owe their physical origins to the existence of supersonic turbulence in the presence of gravity in molecular clouds.
Starting with Larson's original paper, the LWS relation has been generally assumed to be indicative of clouds characterized by turbulent velocity fields. As mentioned earlier cloud simulations which couple gravity and turbulence are capable of generating a scaling between line width and size, although the slope of the relation is not yet sufficiently constrained by either observations or theory. Moreover, the LWS for the diffuse GMC population in M31 is a very weak correlation with  a least squares fit to the relation having a slope of 0.43 but a correlation coefficient of only 0.36. 

In contrast, the masses and sizes of the diffuse GMCs are strongly correlated with a least-squared fitted slope of 2.0 and a correlation coefficient of 0.80. 
The  physical basis for the mass-size relation is perhaps better understood and likely derives from the intrinsic stratification of molecular clouds whose structure can be described by power-law N-pdfs that are steeply falling (i.e., power law index n$>$ 2) with increasing column density. As earlier shown (i.e., Equation 1), a molecular cloud with a pdf that peaks at its outer boundary and rapidly declines inward in a power law or similar fashion will exhibit an average surface density that is only somewhat higher (about a factor of 1.5-2.0) than the value at the outer boundary peak of its pdf, as is observed for Galactic clouds (e.g., Lombardi et al. 2015, Lewis et al. 2022) and demonstrated here for M31 GMCs. As long as the power-law index doesn't vary much within a cloud population, the average surface densities of the GMCs will be roughly constant when measured from similar boundary levels. 
 Although turbulent simulations predict that clouds should have lognormal N-pdfs (e.g., Vazquez-Semadeni 1994, Padoan et al. 1997; Federrath et al. 2010),   it has been suggested that as such clouds evolve in the presence of gravity their pdfs should develop power-law tails and perhaps eventually even full fledged power-law shapes (e.g., Ballesteros-Paredes et al. 2011; Federrath \& Klessen 2013) Thus, it is entirely plausible that both the LWS and mass-size scaling relations can be produced in molecular clouds whose dynamics are governed by both self-gravity and turbulence. More detailed support for these suggestions in models and simulations is yet to be developed.

Finally we note that it is likely that  the diffuse GMCs are not actively forming stars in the present epoch. 
Whether these GMCs have a fundamentally different physical nature than the dense GMCs or are similar objects in different states of evolution, remains to be determined. In the latter case the diffuse clouds could  be either in a very early  pre-star formation evolutionary stage condensing on their way to contracting to form new stars or at an advanced,  post star formation evolutionary stage disrupted by earlier epochs of star formation within them. They could also be ephemeral objects , short-lived and fleeting with little connection to star formation.
In any event,  their existence could explain a discrepancy between the depletion times  
measured for local star forming molecular clouds on scales of $\sim$ 50 pc (i.e., 2 $\times$ 10$^8$ yrs; Lada et al. 2010, 2012) and those measured on kiloparsec scales (i.e. 2 $\times$ 10$^9$ yrs.; Bigiel et al 2008).  
If a significant fraction of the molecular mass on kiloparsec scales in a disk galaxy is in the form of discrete diffuse clouds or spatially diffuse gas, then the discrepancy could be reduced, although by how much is unclear. Unfortunately, we cannot estimate the fraction of spatially diffuse molecular gas from our survey, both because we are only sensitive to detecting discrete clouds and also because our sample may be biased in some unknown way  due to its selection based on a far-infrared catalog of M31. The evolutionary status of the diffuse clouds is an interesting issue given the fact that not being gravitationally bound does suggest relatively short lifetimes for these objects. Local clouds like the Pipe Nebula and the California Molecular Cloud, though presently bound, could be examples of such objects and both appear to be clouds in early evolutionary states. However, it is generally believed that feedback from star formation severely curtails the lifetimes of GMCs once the star formation process takes hold. Unfortunately, there is as yet no consensus regarding the observable properties of such disrupted clouds, and no example of such an object yet unambiguously identified. So it is unclear if the diffuse GMCs in M31 fall into that category.

\section{Summary and Concluding Remarks}

We resolved and identified 162 individual GMCs in \twco(2-1) emission and 85 individual clumps in \thco (2-1) emission within 80 deep SMA fields observed across the disk of the Andromeda galaxy. 
To reduce and largely eliminate contamination from overlapping clouds along the line-of-sight and ensure  that our sample consists of the highest possible quality, uncontaminated measurements of GMCs in M31, we
 focused the bulk of our analysis on a restricted subset of 117 GMCs that were characterized by relatively simple, single component \twco line profiles. We examined Larson's (1981) relations for the 117 surveyed clouds and found:

1--- A strong correlation exists between the masses and sizes of GMCs  in M31 with M$_{\rm GMC}$  $\sim$  R$^{2.37}$. This relation is somewhat steeper than the constant column density relation found by Larson for Milky Way clouds, although it is consistent with recent \twco studies of local Milky Way clouds (Lewis et al 2022).  

2-- The mass-size relation in M31 is characterized by appreciable scatter. This scatter corresponds to a range in the average GMC mass surface density of roughly an order of magnitude with a median value of 48 \msun pc$^{-2}$. Detailed analysis of the scatter shows that it is almost entirely due to the systematic variation in the adopted 3 sigma outer boundaries of the GMCs imposed by corresponding variations in image noise. 

3--Although the sensitivities of the  \twco images span a relatively wide range, our 60 deepest images are sufficiently sensitive to trace CO emission down to 3 sigma boundary surface densities ranging between $\Sigma_0$ $\approx$ 10-22 \msun pc$^{-2}$ (i.e., A$_V$ $\approx$ 0.5 -1.0 mag), near the HI-H$_2$ phase transition boundary. Such depths are  comparable to those achieved in CO surveys of local Milky Way GMCs. For these M31 GMCs, $<\Sigma_{GMC}> =$ 37 $\pm$ 11 \msun pc$^{-2}$, essentially the same as the average value found for local Milky Way GMCs when traced to similar depths.  

4--The ratio of $<\Sigma_{GMC}>/\Sigma_0$ is found to be constant with increasing $\Sigma_0$  suggesting that on sub cloud scales the M31 GMCs are stratified with power-law or otherwise steeply falling surface density pdfs. 

5--- The linewidth-size realation exhibits only a modest correlation between the \twco velocity dispersions and sizes of the M31 GMCs with $\sigma$ $\sim$ R$^{0.35}$ and significant scatter.   Moreover, the \thco observations barely show a very weak correlation between $\sigma$ and size with $\sigma$(\thco) $\sim$ R$^{0.19}$, a value significantly different from that found for \twco emission. These results suggest that the LWS relation may be scale dependent and that the influence of turbulence on cloud dynamics may begin to diminish on larger scales than previously believed.

6---The \twco data indicate that only 43\% of the GMCs in M31 are gravitationally bound against internal motions.  The well known Larson paradigm that GMCs are in dynamical equilibrium with gravity does not hold for most of the GMCs across M31. However, 94\% of the \thco clumps are found to be gravitationally bound.

7--We identify two classes of GMCs within the sample based on the strength and extent of  their \thco emission. Strong \thco emitting (or dense) GMCs make up 38\% of the sample, but account for 73\% of its total mass. These GMCs contain 80\% of the most massive ($> 10^5$ \msun) objects identified in this study. Weak \thco emitting (or diffuse) GMCs constitute the bulk (62\%) of the sample. These GMCs are generally less massive than strong \thco emitting GMCs and span a wider range of mass but only account for 27\% of the total sample mass. 

8--Both classes of GMCs individually exhibit strong mass-size relations with power-law indices near 2, more consistent with Larson's  constant surface density relation than the M31 aggregate relation. However, the dense GMCs have an average surface density a factor of $\sim$ 1.7 higher than that of the diffuse GMCs which likely accounts for the steeper index (2.37)  of the aggregate mass-size relation. Both classes of GMCs exhibit poorly correlated LWS relations with similar slopes that are comparable to that of the aggregate sample. 

9--The great majority (73\%) of dense GMCs are found to be gravitationally bound and the properties of these GMCs appear consistent with the original Larson relations. The dense GMCs contain all the \thco clumps and 38 of 40 of the dust emitting cores in the GMC sample. These GMCs are likely the sites of active star formation in M31. Only 25\% of the diffuse GMCs are found to be bound, suggesting that these weak \thco emitting clouds may be ephemeral objects. Their exact evolutionary status is unclear. 

10---The high fraction of gravitationally unbound GMCs in the diffuse GMC population of M31 deviates from the Larson paradigm of tightly bound GMCs and as a result decouples the mass-size and line width-size relations so that independent physical causes are now required to understand the origin of each of these scaling relations.   

Finally, we have compared the basic physical properties of  the M31 cloud population with the corresponding ones for clouds in the local Milky Way and found the natures of the two cloud populations in these galaxies to be very similar and almost indistinguishable in many respects. This provides strong confirmation of earlier suggestions regarding the similarity of molecular clouds in the two galaxies made over the past decades (e.g., Vogel et al. 1987; Lada et al. 1988; Loinard et al. 1999). Evidently the environmental conditions in the local Milky Way are close enough to those in the M31 disk to produce populations of GMCs that share very similar fundamental properties. This is despite  the significant restructuring and  galaxy-wide disruption of M31 caused by the recent collision between it and M32 (Block et al. 2006; Dierickx et al. 2014). Apparently the resulting impact, initiated some 200 Myr ago, has not managed to produce a GMC population significantly different than that of the local Milky Way, which has not suffered such a catastrophic collision in the same epoch of cosmic history. Evidently the basic physical properties of individual GMCs and perhaps the star formation process within them are quite robust to such global scale disruptions.

\section{Acknowledgements}

We thank the SMA staff for their able assistance in acquiring the observations reported here, particularly during the difficult time of the recent pandemic. We thank Eric Keto for helpful suggestions on an earlier draft of the paper. We acknowledge that  the Submillimeter Array is a joint project between the Smithsonian Astrophysical Observatory and the Academia Sinica Institute of Astronomy and Astrophysics and is funded by the Smithsonian Institution and the Academia Sinica.

\vfill\eject


\begin{thebibliography}{}

\bibitem[Ballesteros-Paredes et al.(2011)]{2011MNRAS.416.1436B} Ballesteros-Paredes, J., V{\'a}zquez-Semadeni, E., Gazol, A., et al.\ 2011, \mnras, 416, 1436. doi:10.1111/j.1365-2966.2011.19141.x

\bibitem[Ballesteros-Paredes et al.(2012)]{2012MNRAS.427.2562B} Ballesteros-Paredes, J., D'Alessio, P., \& Hartmann, L.\ 2012, \mnras, 427, 2562. doi:10.1111/j.1365-2966.2012.22130.x

\bibitem[Beaumont et al.(2012)]{2012MNRAS.423.2579B} Beaumont, C.~N., Goodman, A.~A., Alves, J.~F., et al.\ 2012, \mnras, 423, 2579. doi:10.1111/j.1365-2966.2012.21061.x

\bibitem[Bertoldi \& McKee(1992)]{1992ApJ...395..140B} Bertoldi, F. \& McKee, C.~F.\ 1992, \apj, 395, 140. doi:10.1086/171638

\bibitem[Bhattacharya et al.(2022)]{2022MNRAS.517.2343B} Bhattacharya, S., Arnaboldi, M., Caldwell, N., et al.\ 2022, \mnras, 517, 2343. doi:10.1093/mnras/stac2703


\bibitem[Bigiel et al.(2008)]{2008AJ....136.2846B} Bigiel, F., Leroy, A., Walter, F., et al.\ 2008, \aj, 136, 2846. doi:10.1088/0004-6256/136/6/2846


\bibitem[Blitz \& Williams(1999)]{1999ASIC..540....3B} Blitz, L. \& Williams, J.~P.\ 1999, The Origin of Stars and Planetary Systems, 540, 3

\bibitem[Blitz \& Rosolowsky(2004)]{2004ApJ...612L..29B} Blitz, L. \& Rosolowsky, E.\ 2004, \apjl, 612, L29. doi:10.1086/424661


\bibitem[Block et al.(2006)]{2006Natur.443..832B} Block, D.~L., Bournaud, F., Combes, F., et al.\ 2006, \nat, 443, 832. doi:10.1038/nature05184

\bibitem[Chen et al.(2020)]{2020MNRAS.493..351C} Chen, B.-Q., Li, G.-X., Yuan, H.-B., et al.\ 2020, \mnras, 493, 351. doi:10.1093/mnras/staa235

\bibitem[Colombo et al.(2014)]{2014ApJ...784....3C} Colombo, D., Hughes, A., Schinnerer, E., et al.\ 2014, \apj, 784, 3. 
doi:10.1088/0004-637X/784/1/3

\bibitem[Dame et al.(1993)]{1993ApJ...418..730D} Dame, T.~M., Koper, E., Israel, F.~P., et al.\ 1993, \apj, 418, 730. doi:10.1086/173430

\bibitem[Dame et al.(2001)]{2001ApJ...547..792D} Dame, T.~M., Hartmann, D., \& Thaddeus, P.\ 2001, \apj, 547, 792. doi:10.1086/318388


\bibitem[Dierickx et al.(2014)]{2014ApJ...788L..38D} Dierickx, M., Blecha, L., \& Loeb, A.\ 2014, \apjl, 788, L38. doi:10.1088/2041-8205/788/2/L38

\bibitem[Elia et al.(2022)]{2022ApJ...941..162E} Elia, D., Molinari, S., Schisano, E., et al.\ 2022, \apj, 941, 162. doi:10.3847/1538-4357/aca27d


\bibitem[Elmegreen(1989)]{1989ApJ...338..178E} Elmegreen, B.~G.\ 1989, \apj, 338, 178. doi:10.1086/167192


\bibitem[Elmegreen \& Scalo(2004)]{2004ARA&A..42..211E} Elmegreen, B.~G. \& Scalo, J.\ 2004, \araa, 42, 211. doi:10.1146/annurev.astro.41.011802.094859


\bibitem[Evans(1999)]{1999ARA&A..37..311E} Evans, N.~J.\ 1999, \araa, 37, 311. doi:10.1146/annurev.astro.37.1.311

\bibitem[Evans et al.(2021)]{2021ApJ...920..126E} Evans, N.~J., Heyer, M., Miville-Desch{\^e}nes, M.-A., et al.\ 2021, \apj, 920, 126. doi:10.3847/1538-4357/ac1425


\bibitem[Faesi et al.(2018)]{2018ApJ...857...19F} Faesi, C.~M., Lada, C.~J., \& Forbrich, J.\ 2018, \apj, 857, 19. doi:10.3847/1538-4357/aaad60

\bibitem[Federrath \& Klessen(2013)]{2013ApJ...763...51F} Federrath, C. \& Klessen, R.~S.\ 2013, \apj, 763, 51. doi:10.1088/0004-637X/763/1/51

\bibitem[Federrath et al.(2010)]{2010A&A...512A..81F} Federrath, C., Roman-Duval, J., Klessen, R.~S., et al.\ 2010, \aap, 512, A81. doi:10.1051/0004-6361/200912437


\bibitem[Forbrich et al.(2020)]{2020ApJ...890...42F} Forbrich, J., Lada, C.~J., Viaene, S., et al.\ 2020, \apj, 890, 42. doi:10.3847/1538-4357/ab68de

\bibitem[Forbrich et al.(2023)]{2023MNRAS.525.5565F} Forbrich, J., Lada, C.~J., Pety, J., et al.\ 2023, \mnras, 525, 5565. doi:10.1093/mnras/stad2600


\bibitem[Fritz et al.(2012)]{2012A&A...546A..34F} Fritz, J., Gentile, G., Smith, M.~W.~L., et al.\ 2012, \aap, 546, A34. doi:10.1051/0004-6361/201118619


\bibitem[Garc{\'\i}a et al.(2014)]{2014ApJS..212....2G} Garc{\'\i}a, P., Bronfman, L., Nyman, L.-{\r{A}}., et al.\ 2014, \apjs, 212, 2. doi:10.1088/0067-0049/212/1/2

\bibitem[Gordon et al.(2006)]{2006ApJ...638L..87G} Gordon, K.~D., Bailin, J., Engelbracht, C.~W., et al.\ 2006, \apjl, 638, L87. doi:10.1086/501046


\bibitem[Gratier et al.(2012)]{2012A&A...542A.108G} Gratier, P., Braine, J., Rodriguez-Fernandez, N.~J., et al.\ 2012, \aap, 542, A108. doi:10.1051/0004-6361/201116612


\bibitem[Hacar et al.(2013)]{2013A&A...554A..55H} Hacar, A., Tafalla, M., Kauffmann, J., et al.\ 2013, \aap, 554, A55. doi:10.1051/0004-6361/201220090

\bibitem[Hacar et al.(2018)]{2018A&A...610A..77H} Hacar, A., Tafalla, M., Forbrich, J., et al.\ 2018, \aap, 610, A77. doi:10.1051/0004-6361/201731894

\bibitem[Heiderman et al.(2010)]{2010ApJ...723.1019H} Heiderman, A., Evans, N.~J., Allen, L.~E., et al.\ 2010, \apj, 723, 1019. doi:10.1088/0004-637X/723/2/1019

\bibitem[Heyer \& Dame(2015)]{2015ARA&A..53..583H} Heyer, M. \& Dame, T.~M.\ 2015, \araa, 53, 583. doi:10.1146/annurev-astro-082214-122324

\bibitem[Heyer et al.(2009)]{2009ApJ...699.1092H} Heyer, M., Krawczyk, C., Duval, J., et al.\ 2009, \apj, 699, 1092. doi:10.1088/0004-637X/699/2/1092


\bibitem[Hughes et al.(2013)]{2013ApJ...779...46H} Hughes, A., Meidt, S.~E., Colombo, D., et al.\ 2013, \apj, 779, 46. doi:10.1088/0004-637X/779/1/46


\bibitem[Jackson et al.(2006)]{2006ApJS..163..145J} Jackson, J.~M., Rathborne, J.~M., Shah, R.~Y., et al.\ 2006, \apjs, 163, 145. doi:10.1086/500091


\bibitem[Kafle et al.(2018)]{2018MNRAS.475.4043K} Kafle, P.~R., Sharma, S., Lewis, G.~F., et al.\ 2018, \mnras, 475, 4043. doi:10.1093/mnras/sty082

\bibitem[Kirk et al.(2015)]{2015ApJ...798...58K} Kirk, J.~M., Gear, W.~K., Fritz, J., et al.\ 2015, \apj, 798, 58. doi:10.1088/0004-637X/798/1/58


\bibitem[Kirk et al.(2017)]{2017ApJ...846..144K} Kirk, H., Friesen, R.~K., Pineda, J.~E., et al.\ 2017, \apj, 846, 144. doi:10.3847/1538-4357/aa8631

\bibitem[Koper et al.(1991)]{1991ApJ...383L..11K} Koper, E., Dame, T.~M., Israel, F.~P., et al.\ 1991, \apjl, 383, L11. doi:10.1086/186229


\bibitem[Kudoh \& Basu(2003)]{2003ApJ...595..842K} Kudoh, T. \& Basu, S.\ 2003, \apj, 595, 842. doi:10.1086/377495

\bibitem[Lada \& Dame(2020)]{2020ApJ...898....3L} Lada, C.~J. \& Dame, T.~M.\ 2020, \apj, 898, 3. doi:10.3847/1538-4357/ab9bfb
\bibitem[Lada et al.(2012)]{2012ApJ...745..190L} Lada, C.~J., Forbrich, J., Lombardi, M., et al.\ 2012, \apj, 745, 190. doi:10.1088/0004-637X/745/2/190

\bibitem[Lada et al.(2010)]{2010ApJ...724..687L} Lada, C.~J., Lombardi, M., \& Alves, J.~F.\ 2010, \apj, 724, 687. doi:10.1088/0004-637X/724/1/687

\bibitem[Lada et al.(1988)]{1988ApJ...328..143L} Lada, C.~J., Margulis, M., Sofue, Y., et al.\ 1988, \apj, 328, 143. doi:10.1086/166275

\bibitem[Lada et al.(2008)]{2008ApJ...672..410L} Lada, C.~J., Muench, A.~A., Rathborne, J., et al.\ 2008, \apj, 672, 410. doi:10.1086/523837


\bibitem[Larson(1981)]{1981MNRAS.194..809L} Larson, R.~B.\ 1981, \mnras, 194, 809. doi:10.1093/mnras/194.4.809

\bibitem[Lewis et al.(2022)]{2022ApJ...931....9L} Lewis, J.~A., Lada, C.~J., \& Dame, T.~M.\ 2022, \apj, 931, 9. doi:10.3847/1538-4357/ac5d58

\bibitem[Loinard et al.(1999)]{1999A&A...351.1087L} Loinard, L., Dame, T.~M., Heyer, M.~H., et al.\ 1999, \aap, 351, 1087

\bibitem[Lombardi et al.(2015)]{2015A&A...576L...1L} Lombardi, M., Alves, J., \& Lada, C.~J.\ 2015, \aap, 576, L1. doi:10.1051/0004-6361/201525650

\bibitem[Lombardi et al.(2010)]{2010A&A...512A..67L} Lombardi, M., Lada, C.~J., \& Alves, J.\ 2010, \aap, 512, A67. doi:10.1051/0004-6361/200912670


\bibitem[Maloney(1990)]{1990ApJ...348L...9M} Maloney, P.\ 1990, \apjl, 348, L9. doi:10.1086/185618

\bibitem[Maloney(1988)]{1988ApJ...334..761M} Maloney, P.\ 1988, \apj, 334, 761. doi:10.1086/166876

\bibitem[McConnachie et al.(2005)]{2005MNRAS.356..979M} McConnachie, A.~W., Irwin, M.~J., Ferguson, A.~M.~N., et al.\ 2005, \mnras, 356, 979. doi:10.1111/j.1365-2966.2004.08514.x


\bibitem[Miville-Desch{\^e}nes et al.(2017)]{2017ApJ...834...57M} Miville-Desch{\^e}nes, M.-A., Murray, N., \& Lee, E.~J.\ 2017, \apj, 834, 57

\bibitem[Neumann et al.(2023)]{2023MNRAS.521.3348N} Neumann, L., Gallagher, M.~J., Bigiel, F., et al.\ 2023, \mnras, 521, 3348. doi:10.1093/mnras/stad424


\bibitem[Padoan et al.(1997)]{1997ApJ...474..730P} Padoan, P., Jones, B.~J.~T., \& Nordlund, {\r{A}}. P.\ 1997, \apj, 474, 730. doi:10.1086/303482


\bibitem[Passot et al.(1988)]{1988A&A...197..228P} Passot, T., Pouquet, A., \& Woodward, P.\ 1988, \aap, 197, 228


\bibitem[Rice et al.(2016)]{2016ApJ...822...52R} Rice, T.~S., Goodman, A.~A., Bergin, E.~A., et al.\ 2016, \apj, 822, 52


\bibitem[Roman-Duval et al.(2016)]{2016ApJ...818..144R} Roman-Duval, J., Heyer, M., Brunt, C.~M., et al.\ 2016, \apj, 818, 144. doi:10.3847/0004-637X/818/2/144

\bibitem[Roman-Duval et al.(2010)]{2010ApJ...723..492R} Roman-Duval, J., Jackson, J.~M., Heyer, M., et al.\ 2010, \apj, 723, 492

\bibitem[Rosolowsky(2007)]{2007ApJ...654..240R} Rosolowsky, E.\ 2007, \apj, 654, 240. doi:10.1086/509249

\bibitem[Sanders et al.(2012)]{2012ApJ...758..133S} Sanders, N.~E., Caldwell, N., McDowell, J., et al.\ 2012, \apj, 758, 133. doi:10.1088/0004-637X/758/2/133


\bibitem[Solomon et al.(1987)]{1987ApJ...319..730S} Solomon, P.~M., Rivolo, A.~R., Barrett, J., et al.\ 1987, \apj, 319, 730. doi:10.1086/165493

\bibitem[Sun et al.(2020)]{2020ApJ...892..148S} Sun, J., Leroy, A.~K., Ostriker, E.~C., et al.\ 2020, \apj, 892, 148. doi:10.3847/1538-4357/ab781c

\bibitem[Tabatabaei \& Berkhuijsen(2010)]{2010A&A...517A..77T} Tabatabaei, F.~S. \& Berkhuijsen, E.~M.\ 2010, \aap, 517, A77. doi:10.1051/0004-6361/200913593

\bibitem[Vazquez-Semadeni(1994)]{1994ApJ...423..681V} Vazquez-Semadeni, E.\ 1994, \apj, 423, 681. doi:10.1086/173847


\bibitem[V{\'a}zquez-Semadeni et al.(1997)]{VBR97} V{\'a}zquez-Semadeni, E., Ballesteros-Paredes, J., \& Rodr{\'\i}guez, L.~F.\ 1997, \apj, 474, 292. doi:10.1086/303432


\bibitem[Viaene et al.(2021)]{v21} Viaene, S., Forbrich, J., Lada, C.~J., et al.\ 2021, \apj, 912, 68. doi:10.3847/1538-4357/abe629

\bibitem[Vogel et al.(1987)]{1987ApJ...321L.145V} Vogel, S.~N., Boulanger, F., \& Ball, R.\ 1987, \apjl, 321, L145. doi:10.1086/185022


\bibitem[Watkins et al.(2019)]{2019ApJ...873..118W} Watkins, L.~L., van der Marel, R.~P., Sohn, S.~T., et al.\ 2019, \apj, 873, 118. doi:10.3847/1538-4357/ab089f

\bibitem[Wenger et al.(2019)]{2019ApJ...887..114W} Wenger, T.~V., Balser, D.~S., Anderson, L.~D., et al.\ 2019, \apj, 887, 114. doi:10.3847/1538-4357/ab53d3


\end{thebibliography}
\end{document}